\documentclass[twocolumn,aps,pra,floatfix,superscriptaddress,notitlepage]{revtex4-1}
\usepackage[utf8]{inputenc}
\usepackage[OT1]{fontenc}
\usepackage{slashed}
\usepackage{amsmath}
\usepackage{amsfonts}
\usepackage{amssymb}
\usepackage{bm}
\usepackage{graphicx}
\usepackage[left=2cm,right=2cm,top=2cm,bottom=2cm]{geometry}
\usepackage{longtable,booktabs}
\usepackage{afterpage}
\allowdisplaybreaks

\usepackage{dcolumn}
\usepackage{siunitx}
\usepackage{multirow}
\usepackage{bigdelim}

\usepackage{color}
\definecolor{BLUE}{rgb}{0.0,0.0,1.0}

\usepackage{soul}
\soulregister\cite7
\soulregister\ref7

\newcommand{\veps}{\varepsilon}
\newcommand{\balpha}{\bm{\alpha}}

\newcommand{\bsigma}{\bm{\sigma}}

\newcommand{\bx}{\bm{x}}
\newcommand{\by}{\bm{y}}
\newcommand{\bz}{\bm{z}}
\newcommand{\bp}{\bm{p}}

\newcommand{\bq}{\bm{q}}

\newcommand{\bnx}{\bm{\hat x}}
\newcommand{\bnp}{\bm{\hat p}}

\newcommand{\be}{\begin{eqnarray}}
\newcommand{\ee}{\end{eqnarray}}

\newcommand{\dvec}[2]{{\begin{pmatrix} #1 \\ #2 \end{pmatrix}}}

\begin{document}

\title{Convergence-acceleration approach to partial-wave expansion \\ of two-electron self-energy contributions to the Lamb shift}

\author{A.~V.~Malyshev}
\affiliation{Department of Physics, St.~Petersburg State University, Universitetskaya 7-9, 199034 St.~Petersburg, Russia  
\looseness=-1}
\affiliation{Petersburg Nuclear Physics Institute named by B.P. Konstantinov of National Research Center ``Kurchatov Institute'', Orlova roscha 1, 188300 Gatchina, Leningrad region, Russia}

\author{E.~A.~Prokhorchuk}
\affiliation{Department of Physics, St.~Petersburg State University, Universitetskaya 7-9, 199034 St.~Petersburg, Russia  
\looseness=-1}

\author{V.~M.~Shabaev}
\affiliation{Department of Physics, St.~Petersburg State University, Universitetskaya 7-9, 199034 St.~Petersburg, Russia  
\looseness=-1}
\affiliation{Petersburg Nuclear Physics Institute named by B.P. Konstantinov of National Research Center ``Kurchatov Institute'', Orlova roscha 1, 188300 Gatchina, Leningrad region, Russia}


\begin{abstract}

Methods of bound-state QED that treat the self-energy contributions to the Lamb shift within the partial-wave expansion usually face the problem of slow convergence of the latter. Inspired by an approach formulated in [J.~Sapirstein and K.~T.~Cheng, Phys. Rev. A {\bf 108}, 042804 (2023)], we propose a modification of the standard procedure for calculating the contributions of two-electron self-energy diagrams. The performance of the method is studied by evaluating the corresponding corrections to the binding energies of He-like ions and by comparing the obtained results with the state-of-the-art values available in the literature: our calculations involving a much smaller number of partial waves show an improvement in accuracy.

\end{abstract}


\maketitle


\section{Introduction \label{sec:0}}

Bound-state quantum electrodynamics (QED) provides a consistent description of various atomic properties, e.g., energy spectra, $g$ factors, hyperfine splittings, etc., by means of perturbation series, conveniently represented by Feynman diagrams. The corresponding calculations are often to be carried out nonperturbatively in the parameter $\alpha Z$, where $\alpha$ is the fine-structure constant and $Z$ is the nuclear-charge number. This is obviously crucial for highly charged ions, where $\alpha Z$ is close to unity~\cite{Sapirstein:2008:25, Shabaev:2018:60, Indelicato:2019:232001}. However, the all-order (in $\alpha Z$) calculations find their application even for lightest H-like atoms, see, e.g., Refs.~\cite{Yerokhin:2019:1800324, Tiesinga:2021:025010} and references therein.

\begin{figure}
\begin{center}
\includegraphics[height=0.375\columnwidth]{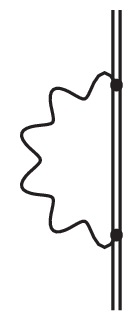}
\caption{\label{fig:se}
First-order self-energy diagram. The double line denotes the electron propagator in the binding potential~$V(x)$; for point nucleus $V_{\rm C}(x)=-\alpha Z/x$. The wavy line corresponds to the photon propagator. The mass-counterterm diagram is omitted.}
\end{center}
\end{figure}

To be more specific and illustrate the issues under discussion, let us consider the first-order self-energy (SE) contribution to an energy level~$|a\rangle$. The corresponding Feynman diagram is shown in Fig.~\ref{fig:se}; the related mass-counterterm diagram is omitted. The energy shift due to the SE diagram is given by the real part of the expression
\begin{align}
\label{eq:SE}
E_{\rm 1} &=\, 2i\alpha \int_{-\infty}^{\infty} \!\! d\omega \int \! d\bx_1 d\bx_2 \, 
D_{\mu\nu}(\omega,\bx_{12}) \psi^\dagger_a(\bx_1) \alpha^{\mu} \nonumber \\
& \!\!\! \times G(\veps_a-\omega,\bx_1,\bx_2) \alpha^{\nu} \psi_a(\bx_2)
- \delta m \int \! d \bx \, \psi^\dagger_a(\bx) \beta \psi_a(\bx) \, ,
\end{align}
where $\psi_a$ and $\veps_a$ are the Dirac wave function and energy of the state $|a\rangle$, $\alpha^{\mu}=(1,\balpha)$, $\balpha$ and $\beta$ are the Dirac matrices, $\bx_{12}=\bx_1-\bx_2$, $D_{\mu\nu}(\omega,\bx_{12})$ is the photon propagator, $G(E,\bx_1,\bx_2)$ is the bound-electron Green's function, i.e., the electron propagator in the local binding potential $V(\bx)$, and $\delta m$ is the mass counterterm. We note that one- and two-electron vacuum-polarization diagrams, which contribute to the same orders of QED perturbation theory as the SE ones discussed below, are beyond the scope of the present work.

The calculations of the first-order SE correction to all orders in $\alpha Z$ have a long history and began with Refs.~\cite{Desiderio:1971:1267, Mohr:1974:26, Mohr:1974:52}. In view of the importance of the SE contribution, many methods for its evaluation have been proposed in the literature~\cite{Mohr:1974:26, Mohr:1974:52, Snyderman:1991:43, Blundell:1991:R1427, Blundell:1992:3762, Mohr:1992:4421, Indelicato:1992:172, Cheng:1993:1817, Persson:1993:125, Quiney:1993:132, Quiney:1994:L299, Labzowsky:1997:177, Indelicato:1998:165, Jentschura:1999:53, Yerokhin:1999:800, LeBigot:2001:052508, Hedendahl:2012:012514}. These methods differ, in particular, in the way they handle the ultraviolet (UV) divergences. In this work, a variation of the potential-expansion (PE) approach, introduced for the SE diagrams in Ref.~\cite{Snyderman:1991:43}, is considered. Within the PE methods~\cite{Blundell:1991:R1427, Blundell:1992:3762, Cheng:1993:1817, Yerokhin:1999:800}, the UV-divergent terms are separated out by expanding the bound-electron Green's function~$G$ in terms of the potential~$V$, $G=\sum_{i=0}^\infty G^{(i)}$, where the index $i$ denotes the power of $V$, i.e., $G^{(0)}$ is the free-electron Green's function, and $G^{(i+1)}=G^{(i)}VG^{(0)}$ (the vertex-coordinate integration is implied in such shorthand notations). The UV-divergent terms are calculated in momentum space after a renormalization. For the SE diagram in Fig.~\ref{fig:se}, the UV divergences are associated with the terms~$G^{(0)}$ and $G^{(1)}$. The approaches to treat the SE diagram also differ in the following aspect. The closed analytical form of the electron propagator in a spherically symmetric potential~$V(x)\neq 0$, where $x=|\bx|$, is currently unknown. For this reason, the bound-electron Green's function is inevitably represented in atomic calculations by an infinite sum of terms corresponding to the different values of relativistic angular quantum number~$\kappa=(-1)^{j+l+1/2}(j+1/2)$, where $l$ and $j$ are the orbital and total angular momenta, respectively. For instance, in the original approach developed by Mohr~\cite{Mohr:1974:26}, the partial-wave summation is performed numerically in the integrand of Eq.~(\ref{eq:SE}), until the desired accuracy is achieved. In contrast, the PE methods usually treat the partial waves step by step, integrating each term independently and then analyzing the convergence of the partial-wave expansion, which may be quite slow in some cases. The ``pros'' and ``cons'' of the PE methods have been discussed many times, so for details we refer the reader, e.g., to Ref.~\cite{Yerokhin:2005:042502}. As an advantage of the PE methods, we just note that they can be readily generalized for calculating the contributions of higher-order Feynman diagrams. For a review of the issues concerning various applications of the Green's function and different ways to represent it in QED calculations, see, e.g., Ref.~\cite{Yerokhin:2020:800} and references therein. 

A number of modifications have been proposed to accelerate the convergence of partial-wave expansion in the PE methods for the SE diagram in Fig.~\ref{fig:se}. Some of them exploit the idea, originally formulated by Mohr in Ref.~\cite{Mohr:1974:26}, that the dominant contribution to an expression containing the free-electron Green's function $G^{(0)}(E,\bx_1,\bx_2)$ comes from the region where $\bx_1 \approx \bx_2$ and, therefore, the result will not change significantly, if one replaces $V(x_1)G^{(0)}(E,\bx_1,\bx_2)$ with $G^{(0)}(E,\bx_1,\bx_2)V(x_2)$ or vica versa. This trick was employed, e.g., in Ref.~\cite{Yerokhin:2005:042502} in order to obtain an approximation for the PE remainder~$G^{(2+)}\equiv G-G^{(0)}-G^{(1)}$, which corresponds to the terms with two or more potentials. The key point is that, on the one hand, the constructed approximate expression can be calculated numerically with high precision and, on the other hand, it has a partial-wave expansion which can be subtracted from that of the initial PE method, resulting in a better convergence of the difference. Technically, the modification represents an identity transformation of the original PE expression. One adds and subtracts the same quantity, but evaluates it in different ways, which ultimately determines an improvement in accuracy. For instance, in the modification of Ref.~\cite{Yerokhin:2005:042502}, the subtracted contribution, when evaluated without recourse to the partial-wave expansion, is treated in coordinate space using the closed-form expression (without the partial-wave expansion) for the free-electron Green's function~\cite{Mohr:1974:26}. A similar approach has been proposed recently in Ref.~\cite{Sapirstein:2023:042804}, where the discussed trick was employed to construct an approximation for the PE term~$G^{(2)}$ containing two potentials. With some simple reasoning, it is possible to derive the closed formula for this subtraction as well. However, in this case, in contrast to Ref.~\cite{Yerokhin:2005:042502}, the corresponding expression is calculated in momentum space. 

\begin{figure}
\begin{center}
\includegraphics[height=0.45\columnwidth]{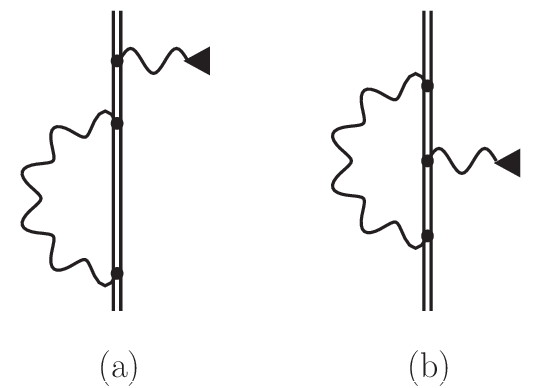}
\caption{\label{fig:se_ext}
Self-energy diagrams in the presence of an additional perturbing field~$\delta V$. Wavy line ended with a black triangle denotes the interaction with~$\delta V$. The other notations are as in Fig.~\ref{fig:se}.}
\end{center}
\end{figure}

The success of modifications proposed in Refs.~\cite{Yerokhin:2005:042502, Sapirstein:2023:042804} supports a natural conjecture that the slow convergence of the partial-wave expansion is mainly related to the next-to-divergent terms of the PE. The difficulty is that in coordinate space the term~$G^{(2)}$, treated without any of the approximations discussed above, has no closed-form representation, while in momentum space it leads to a multidimensional integral, whose accurate evaluation constitutes a challenging problem. 

Nevertheless, such a subtraction of $G^{(2)}$ (without an approximation) has been implemented in some form in Refs.~\cite{Artemyev:2007:173004, Artemyev:2013:032518} for approaches that use a spectral decomposition within a finite-basis set to represent the free- and bound-electron Green's functions~\cite{Yerokhin:2020:800}. As is well known, the number of basis functions, e.g., B-splines~\cite{Sapirstein:1996:5213, splines:DKB}, required for an adequate representation of electron propagators grows rapidly with increasing $|\kappa|$, making the calculations very time-consuming for large $|\kappa|$. In this connection, the calculations utilizing the finite-basis-set representation of Green's functions are typically restricted to fewer partial waves than the other PE methods~\cite{Yerokhin:2020:800, Artemyev:2013:032518}. To improve the convergence in the case of propagators represented in this way, the term~$G^{(2)}$, evaluated within the same finite-basis-set approach, is additionally subtracted from $G^{(2+)}$, turning it into $G^{(3+)}$~\cite{Artemyev:2007:173004, Artemyev:2013:032518}. In view of the aforementioned difficulties, it has been proposed to estimate the total value of the subtraction also in coordinate space within the same partial-wave expansion, but to extend the calculations to the larger values of $|\kappa|$~\cite{Artemyev:2007:173004, Artemyev:2013:032518}, and then add it to the final result. This becomes possible, since the finite-basis-set representation of the free-electron Green's function is not used in this case. In fact, this modification transfers the slow-convergence problem from the evaluation of the total SE correction to the calculation of its two-potential part.

\begin{figure}
\begin{center}
\includegraphics[height=0.45\columnwidth]{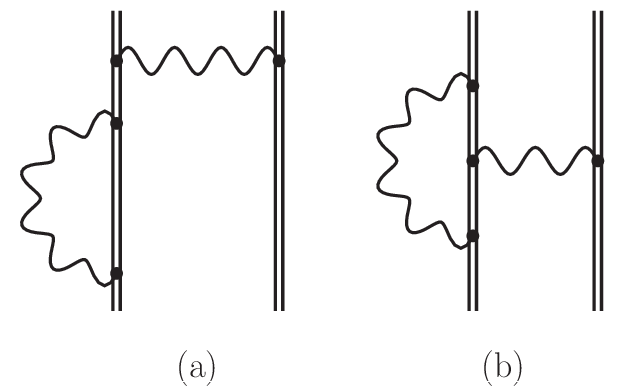}
\caption{\label{fig:scrse}
Two-electron self-energy diagrams. The notations are the same as in Fig.~\ref{fig:se}.}
\end{center}
\end{figure}

It is worth noting that a separate treatment of next-to-leading PE terms to improve the convergence has been realized for the SE corrections to the bound-electron $g$ factor~\cite{Yerokhin:2004:052503, Persson:1997:R2499} and hyperfine splitting (HFS)~\cite{Yerokhin:2010:012502}. The relevant Feynman diagrams are shown in Fig.~\ref{fig:se_ext}, where the potential $\delta V$ perturbing the ordinary SE diagram is given by a constant external magnetic field in the case of the $g$ factor and by the magnetic field of the nucleus in the case of the HFS. In the diagram in Fig.~\ref{fig:se_ext}(a), the potential $\delta V$ does not enter the SE loop. Therefore, the methods developed for the first-order SE correction in Eq.~(\ref{eq:SE}) can be adapted for the corresponding contribution. For the vertex diagram in Fig.~\ref{fig:se_ext}(b), this is not the case. Here, the term $G^{(0)} \delta V G^{(0)}$ is UV-divergent, and, to accelerate the convergence of the partial-wave expansion, the next term of the PE, $G^{(0)} \delta V G^{(1)} + G^{(1)} \delta V G^{(0)}$, has to be additionally subtracted. For the bound-electron $g$ factor, due to a special form of the potential $\delta V$ in momentum space, the dimensionality of the corresponding integral for the subtraction is considerably reduced, and its evaluation turns out to be similar to that for the renormalized term $G^{(1)}$ in Eq.~(\ref{eq:SE})~\cite{Yerokhin:2004:052503}. For this reason, this method has become a standard practice in the bound-electron $g$-factor calculations. For the HFS, such a reduction does not occur. The expression for the subtraction was derived in momentum space for the point-nucleus Coulomb potential~$V_{\rm C}(x)=-\alpha Z/x$, it contains seven integrations, and its accurate evaluation has required a lot of computing time and the use of quadruple-precision arithmetic~\cite{Yerokhin:2010:012502}. Such calculations are unique, and they are hard to be serialized.

The two-electron SE diagrams are shown in Fig.~\ref{fig:scrse}. An accurate evaluation of these contributions is essential for a proper QED description of few-electron systems in a wide range of $Z$~\cite{Artemyev:2005:062104, Sapirstein:2011:012504, Malyshev:2021:183001, Yerokhin:2022:022815, Malyshev:2023:042806}. All the methods developed for the first-order SE diagram in Fig.~\ref{fig:se} can be carried over to the case of the diagram in Fig.~\ref{fig:scrse}(a). On the other hand, as far as we know, there are no convergence-acceleration schemes reported in the literature for the vertex diagram in Fig.~\ref{fig:scrse}(b). The truncation of the partial-wave expansion for the corresponding contribution is the main source of the numerical uncertainty~\cite{Yerokhin:2022:022815}. The modification described in Ref.~\cite{Yerokhin:2005:042502} can, in principle, be extended for computations of the vertex diagram. However, this would require substantial changes to the available and well-established numerical codes, which is a laborious task. In this regard, an extension of the approach suggested in Ref.~\cite{Sapirstein:2023:042804} seems to be more straightforward. For this reason, the aim of the present work is to describe the method, capable of improving the partial-wave-expansion convergence in the case of the two-electron SE diagrams, and to study its performance. The main focus will be on the vertex diagram in Fig.~\ref{fig:scrse}(b).

The paper is organized as follows. In Sec.~\ref{sec:1}, the basic formulas for the two-electron SE diagrams are briefly reviewed and the standard PE approach outlined in Refs.~\cite{Yerokhin:1999:800, Yerokhin:1999:3522} is discussed. Sec.~\ref{sec:2} is devoted to the description of the convergence-acceleration method as applied to the two-electron SE diagrams. In Sec.~\ref{sec:3}, we perform test calculations for He-like ions, analyze the convergence of partial-wave expansions, and compare the results obtained with the those available in the literature. The partial-wave expansions of photon and electron propagators are discussed in Appendix~\ref{sec:app:1}. Some details of the momentum-space calculations required to improve the convergence are given in Appendices~\ref{sec:app:2} and~\ref{sec:app:3}.

Relativistic units ($\hbar=1$ and $c=1$) and Heaviside charge unit ($e^2=4\pi\alpha$, where $e<0$ is the electron charge) are used throughout the paper. 


\section{Basic formulas \label{sec:1}}

The formal expressions corresponding to the two-electron SE diagrams in Fig.~\ref{fig:scrse} can be readily obtained, e.g., within the two-times Green's function method~\cite{TTGF}. The mass-counterterm diagrams are not not explicitly specified in Fig.~\ref{fig:scrse}, but we properly take them into account during a renormalization. For the sake of simplicity, we assume that unperturbed wave functions are represented by two-electron Slater determinants,
\begin{align}
\label{eq:u_2el}
u(\bx_1,\bx_2) = \frac{1}{\sqrt{2}} \sum_P (-1)^P \psi_{Pa}(\bx_1)\psi_{Pb}(\bx_2) \, , 
\end{align} 
where $P$ is the permutation operator, $(-1)^P$ is its sign, and $\psi$ stands for the solutions of the one-electron Dirac 
equation in some local binding potential~$V$,
\begin{equation}
\label{eq:DirEq}
 \left[ \balpha \cdot \bp + \beta m + V \right] \psi_n = \veps_n \psi_n \, .
\end{equation}
The transition to the general case of many-determinant wave functions is straightforward. Moreover, although we consider only the simplest case of perturbation theory for a single level, everything discussed below can be applied to mixing states~\cite{TTGF} as well. 

First, let us introduce some basic operators and formulas. The interelectronic-interaction operator $I(\omega)$ is defined by
\begin{align}
\label{eq:I}
I(\omega,\bx_1,\bx_2) = e^2 \alpha^\mu \alpha^\nu D_{\mu \nu}(\omega,\bx_{12}) \, .
\end{align}
Representing the electron Green's function as
\begin{align}
\label{eq:G}
G(E,\bx_1,\bx_2) = \sum_n \frac{\psi_n(\bx_1) \psi^\dagger_n(\bx_2)}{E-u\veps_n} \, ,
\end{align}
where $u=1-i0$ and the sum over $n$ is extended over the complete Dirac spectrum, one can express the matrix element of the unrenormalized one-loop SE operator~$\Sigma(\veps)$ as follows,
\begin{align}
\label{eq:Sigma}
\langle a |\Sigma(\veps)|b\rangle = \frac{i}{2\pi} \int_{-\infty}^{\infty} \!\! d\omega
\sum_n \frac{\langle a n | I(\omega)|n b \rangle}{\veps-\omega - u\veps_n} \, .
\end{align}
Employing Eq.~(\ref{eq:G}) and the orthogonality of the functions $\psi_n$, the following identities can be obtained:
\begin{widetext}
\begin{align}
\label{eq:GG}
&\int \! d\by \, G(E,\bx_1,\by)G(E,\by,\bx_2) = -\frac{\partial }{\partial E} G(E,\bx_1,\bx_2) \, ,\\
\label{eq:GGG}
&\int \! d\by d\bz \, G(E,\bx_1,\by)G(E,\by,\bz)G(E,\bz,\bx_2)    
= \frac{1}{2} \frac{\partial^2 }{\partial E^2} G(E,\bx_1,\bx_2) \, .
\end{align}
\end{widetext}
The same expressions are valid for the free-electron Green's function~$G^{(0)}$ as well. We also use the notations: $\Delta_{a,b}=\veps_a-\veps_b$, $I'(\omega) = \partial I(\omega)/\partial \omega$, and $\Sigma'(\veps)=\partial \Sigma(\veps)/\partial \veps$.

The contribution of the diagram in Fig.~\ref{fig:scrse}(a) is naturally divided into the reducible (``red'') and irreducible (``irr'') parts. The reducible part comes from the intermediate two-electron states whose energy coincides with the unperturbed energy $E^{(0)}=\veps_a + \veps_b$, while the irreducible part represents the remainder. In turn, the reducible contribution is conveniently represented by a sum of two terms, defined according to whether the derivative with respect to an energy parameter acts on the operators $I$ or $\Sigma$. Following Ref.~\cite{Yerokhin:1999:3522}, we denote these terms by A and B. The derivation of the formal expression for the vertex (``vert'') diagram in Fig.~\ref{fig:scrse}(b) does not require any additional separation. Therefore, the total two-electron SE contribution is given by
\begin{align}
\label{eq:E_tot}
E_{\rm 2}=E_{\rm irr} + E^{\rm A}_{\rm red} + E^{\rm B}_{\rm red} + E_{\rm vert} \, , 
\end{align}
where
\begin{widetext}
\begin{align}
\label{eq:E_irr}
&E_{\rm irr} = 2 \sum_{PQ} (-1)^{P+Q} \sum_{n\neq Pa} 
\langle Pa | \Sigma (\veps_{Pa}) | n \rangle 
\frac{ \langle n Pb | I (\Delta_{Qb,Pb})|QaQb \rangle}{\veps_{Pa} - \veps_n} \, , \\
\label{eq:E_redA}
&E^{\rm A}_{\rm red} = \sum_{PQ} (-1)^{P+Q} \langle Pa | \Sigma (\veps_{Pa}) | Pa \rangle
\langle Pa Pb | I'(\Delta_{Qb,Pb}) | Qa Qb \rangle  \, , \\
\label{eq:E_redB}
&E^{\rm B}_{\rm red} = \sum_{PQ} (-1)^{P+Q} \langle Pa | \Sigma' (\veps_{Pa}) | Pa \rangle
\langle Pa Pb | I(\Delta_{Qb,Pb}) | Qa Qb \rangle  \, , \\
\label{eq:E_vert}
&E_{\rm vert} =\sum_{PQ} (-1)^{P+Q} \frac{i}{2\pi}  \int_{-\infty}^{\infty} \!\! d\omega
\sum_{n_1n_2} \frac{\langle Pa \,n_2 |I(\omega)|n_1Qa\rangle \langle n_1 Pb | I(\Delta_{Qb,Pb})|n_2Qb\rangle}
{(\veps_{Pa}-\omega-u\veps_{n_1})(\veps_{Qa}-\omega-u\veps_{n_2})} \, .
\end{align}
\end{widetext}
The evaluation of the terms $E_{\rm irr}$ and $E^{\rm A}_{\rm red}$ is similar to the calculations of the one-electron SE contribution. The terms $E^{\rm B}_{\rm red}$ and $E_{\rm vert}$ are infrared (IR) divergent, when considered separately. For these reasons, in the numerical calculations, we group the individual two-electron SE contributions as follows,
\begin{align}
\label{eq:E_tot_2}
E_{\rm 2}=E_{\rm irA} + E_{\rm vr} \, , 
\end{align}
where $E_{\rm irA} = E_{\rm irr} + E^{\rm A}_{\rm red}$ and $E_{\rm vr} = E^{\rm B}_{\rm red} + E_{\rm vert}$. In principle, the IR-divergent terms can be explicitly separated out and regularized  by introducing a finite photon mass~\cite{Yerokhin:2020:800}. However, we prefer to handle the IR divergences numerically by combining together all the relevant terms before performing the $\omega$ integration.

\begin{figure}
\begin{center}
\includegraphics[height=0.35\columnwidth]{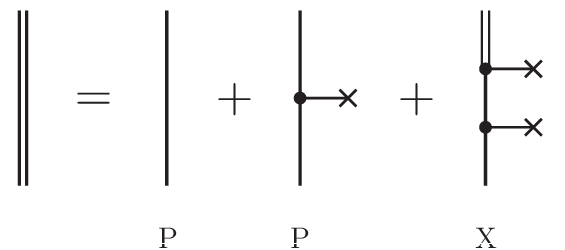}
\caption{\label{fig:calc_1}
Standard calculation scheme for the one-electron SE diagram, which is also suitable for the sum of irreducible and A-reducible terms. The double and single lines represent the bound- and free-electron Green's functions. The line ending with a small cross stands for the interaction with the binding potential~$V$. The photon and external electron lines are omitted. The letters ``P'' and ``X'' denote that the corresponding terms are treated in momentum and coordinate spaces, respectively.}
\end{center}
\end{figure}

The expressions~(\ref{eq:E_irr})-(\ref{eq:E_vert}) suffer from UV divergences. In order to eliminate them, we employ the renormalization procedures worked out in Refs.~\cite{Yerokhin:1999:800, Yerokhin:1999:3522}. Namely, the bound-electron Green's functions are expanded in terms of the binding potential~$V$, the UV-divergent contributions of the PE are separated out and then treated in momentum space, where the divergences are covariantly regularized and explicitly canceled. While in Appendices~\ref{sec:app:2} and \ref{sec:app:3} some relevant formulas are given, we do not focus on the terms requiring the renormalization in the present work and refer the reader to the original studies of the issues~\cite{Yerokhin:1999:800, Yerokhin:1999:3522}. 

For the contribution~$E_{\rm irA}$, the PE decomposition of the Green's function~$G$ necessary for the renormalization is shown schematically in Fig.~\ref{fig:calc_1}. First two terms of the PE, usually referred to as the zero- and one-potential contributions, are separated out. Not to overload the schemes, we draw only the inner parts of the SE loops omitting the photon propagator and external electron lines. Here and below, the letters ``P'' or ``X''  indicate the space, momentum or coordinate, in which the corresponding terms are evaluated. In the PE approaches, the X-space calculations imply the application of a partial-wave expansion. This expansion is truncated at some level with a subsequent extrapolation to infinity. As a rule, this is the main source of the uncertainty. In contrast, the P-contributions are treated without any partial-wave expansion, and, in this sense, they are ``exact''. Their uncertainties are determined by the accuracy with which the multidimensional integrals in momentum space can be evaluated. For instance, the second term in the decomposition in Fig.~\ref{fig:calc_1}, which corresponds to $G^{(1)}=G^{(0)}VG^{(0)}$, contains four integrations that do not cause numerical problems. The third term in Fig.~\ref{fig:calc_1} comprises two or more interactions with the potential~$V$. In literature, it is referred to as the many-potential contribution. In practical calculations, this term, $G^{(2+)}$, can be represented in different ways. For example, it can be obtained by calculating in coordinate space the expressions $G-G^{(0)}-G^{(0)}VG^{(0)}$ or $(G-G^{(0)})VG^{(0)}$. In this work, we evaluate the many-potential term literally as depicted in Fig.~\ref{fig:calc_1}: $G^{(2+)}=GVG^{(0)}VG^{(0)}$. 

\begin{figure}
\begin{center}
\includegraphics[height=0.35\columnwidth]{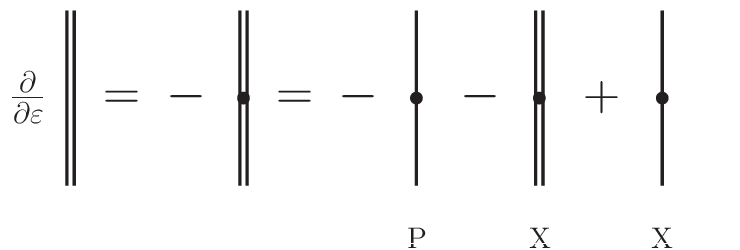}
\caption{\label{fig:calc_2}
Standard calculation scheme for the B-reducible term. The dot on the electron lines stands for the interaction with the identity operator. The other notations are as in Fig.~\ref{fig:calc_1}. The first equality is a graphical representation of Eq.~\ref{eq:GG}.}
\end{center}
\end{figure}

\begin{figure}
\begin{center}
\includegraphics[height=0.35\columnwidth]{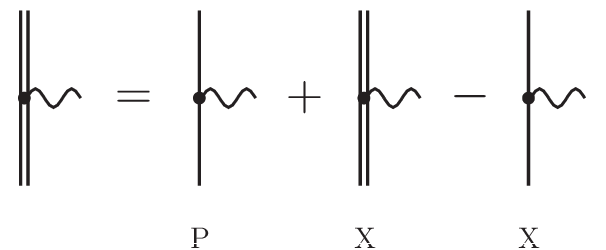}
\caption{\label{fig:calc_3}
Standard calculation scheme for the vertex term. The wavy line represents the virtual photon responsible for the interelectronic interaction. The other notations are as in Fig.~\ref{fig:calc_1}.}
\end{center}
\end{figure}

For the contribution~$E_{\rm vr}$, only the leading term of the PE is UV divergent. The standard calculation schemes for the B-reducible and vertex terms are given in Figs.~\ref{fig:calc_2} and \ref{fig:calc_3}, respectively. According to Eq.~(\ref{eq:GG}), the derivative of the SE operator in Eq.~(\ref{eq:E_redB}) up to a sign can be reduced to the integral of the product of two Green's function interacting via the identity operator. Therefore, from a practical-calculation point of view, the B-reducible term represents a special case of the more complicated vertex contribution, since the identity operator preserves all the angular quantum numbers, while the photon lines can alter them. This is reflected in the fact that the schemes in Figs.~\ref{fig:calc_2} and \ref{fig:calc_3} are similar to each other.

The partial-wave expansion of the SE contributions is based on those of the photon and electron propagators, see Appendix~\ref{sec:app:1}. The sums over $L$ for photons and $\kappa$ for electrons in these expansions are not independent due to the triangular inequalities that the angular momenta should satisfy in all vertices of the SE diagrams. In our calculations, we consider the sums over $\kappa$ as the primary ones and include all relevant values of $L$ for the evaluated electron angular states. This differs, e.g., from Ref.~\cite{Sapirstein:2023:042804}, where the calculations of the individual partial waves were tied to the photon line rather than the electron line. 


The extrapolation of the partial-wave expansion, truncated at some $|\kappa_{\rm max}|$, to infinity is an essential part of the PE methods. In this work, we perform it as follows. The individual contributions for $|\kappa|\leqslant k$ on the electron lines are added to partial sums $S_k$ for $k=1,2,\ldots$. The $|\kappa_{\rm max}|\rightarrow\infty$ limit is defined by polynomial (in $1/k$) least-squares fitting of $S_k$. By trying different orders of the polynomials and different data samples, we obtain an estimate of the uncertainty associated with the extrapolation procedure.


\section{Convergence-acceleration approach \label{sec:2}}

The general idea of the approach discussed below was described in Sec.~\ref{sec:0}: to subtract a slowly-converging part of the partial-wave expansion and calculate it separately using a closed-form momentum-space representation. We emphasize that the contributions considered here do not require any renormalization. In contrast to the P-space contributions shown in Figs.~\ref{fig:calc_1}-\ref{fig:calc_3}, the calculations in momentum space are intended here to improve the behavior of the partial-wave expansions, not to eliminate the UV divergences. 

\begin{figure}
\begin{center}
\includegraphics[height=0.36\columnwidth]{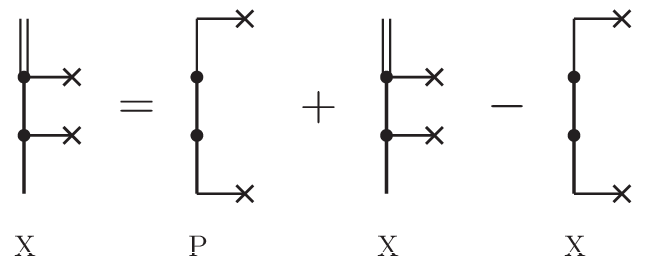}
\caption{\label{fig:calc_4}
Convergence-acceleration scheme for the one-electron self-energy operator~$\Sigma$ proposed in Ref.~\cite{Sapirstein:2023:042804}. The notations are described in Figs.~\ref{fig:calc_1} and \ref{fig:calc_2}.}
\end{center}
\end{figure}

The convergence-acceleration method proposed in Ref.~\cite{Sapirstein:2023:042804} for calculating the SE part of the Lamb shift is shown schematically in Fig.~\ref{fig:calc_4}. This subtraction models the contribution of the next-to-divergent term $G^{(2)}$. The fact that $V(x_1)G^{(0)}(E,\bx_1,\bx_2) \approx G^{(0)}(E,\bx_1,\bx_2)V(x_2)$ justifies moving the potentials~$V$ from the inner electron line to the vertices, where the photon propagator is attached. This approach can be readily generalized to the case of the contribution~$E_{\rm irA}$. The closed-form expression for the P-space term in Fig.~\ref{fig:calc_4} is given in Ref.~\cite{Sapirstein:2023:042804} and discussed in Appendix~\ref{sec:app:2}. In the coordinate space, we evaluate the value of the subtraction as close as possible to the term~$G^{(2+)}$ in order to ensure a numerical cancellation of the slowly-converging contribution.

\begin{figure}
\begin{center}
\includegraphics[height=0.35\columnwidth]{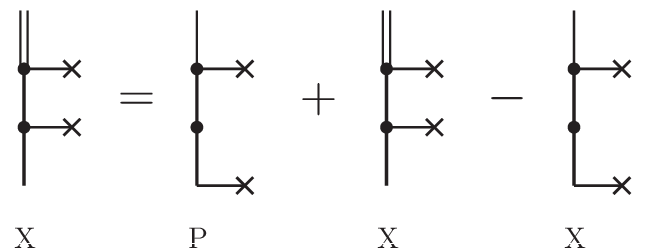}
\caption{\label{fig:calc_5}
Convergence-acceleration scheme for the one-electron self-energy operator~$\Sigma$, which is alternative to that in Fig.~\ref{fig:calc_4}. The notations are described in Figs.~\ref{fig:calc_1} and \ref{fig:calc_2}.}
\end{center}
\end{figure}
\begin{figure*}
\begin{center}
\includegraphics[height=0.36\columnwidth]{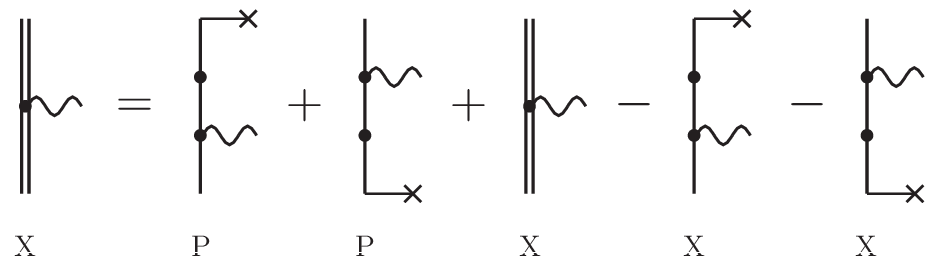}
\caption{\label{fig:calc_6}
Convergence-acceleration scheme for the vertex diagram in Fig.~\ref{fig:scrse}(b). The notations are described in Figs.~\ref{fig:calc_1} and \ref{fig:calc_2}.}
\end{center}
\end{figure*}

In the present work, we also study an alternative scheme to improve the convergence of the many-potential term~$G^{(2+)}$. This scheme is shown in Fig.~\ref{fig:calc_5} and differs from the one in Fig.~\ref{fig:calc_4} in that in this case only one potential in $G^{(2)}$, no matter which of the two, is transposed from the electron line into the diagram vertex. The derivation of the P-space expression for the subtraction is outlined in Appendix~\ref{sec:app:3}. The X-space calculations are similar to each other in both schemes. As it will be seen from the results presented in the next section, the scheme in Fig.~\ref{fig:calc_5} is, in principle, not superior to the original scheme from Ref.~\cite{Sapirstein:2023:042804}. Moreover, the P-space expression in Fig.~\ref{fig:calc_5} turns out to be more complicated than the corresponding term in Fig.~\ref{fig:calc_4}. However, this scheme is an important step toward the convergence-acceleration scheme for the vertex SE diagrams, which are the main focus of the work, and, therefore, it provides a good cross-check.

Let us now discuss the vertex diagram. As noted in the previous section, the UV divergent term corresponds to the leading term of the PE, when both bound-electron Green's functions are replaced with their free-electron counterparts. The next-to-divergent term includes one additional interaction with the potential~$V$. This interaction can be located on one electron line or the other. In the general case, there is no symmetry between the electron lines, therefore, both contributions should be treated on equal footing. Using the same idea, this next-to-divergent term is approximated by transposing the potential to the vertex, where the photon forming the SE loop is attached. The resulting convergence-acceleration scheme is shown in Fig.~\ref{fig:calc_6}. The derivation of the closed-form momentum-space expressions for the subtractions is straightforward, but rather tedious, see the discussion in Appendix~\ref{sec:app:3}. The corresponding formulas involve four-dimensional integrals. The complexity of their evaluation is comparable to that of the P-space term in Fig.~\ref{fig:calc_3}. The additional calculations in the coordinate space do not pose a numerical problem either. Therefore, the proposed method does not significantly complicate the treatment of the two-electron SE diagrams from a technical point of view.

Finally, we note that the scheme presented in Fig.~\ref{fig:calc_6} can also be applied to the calculations of the B-reducible contribution. As noted in the previous section, to this end, the interaction with the photon has to be replaced by the identity operator taken with the opposite sign, see Figs.~\ref{fig:calc_2} and \ref{fig:calc_3}. The states in the matrix element of $\Sigma'$ in Eq.~(\ref{eq:E_redB}) are the same. Therefore, in this case, both subtractions in Fig.~\ref{fig:calc_6} coincide, and the corresponding contributions can be doubled. 


\section{Numerical results and discussions \label{sec:3}}

As a test bed for the approaches discussed above, we chose the simplest system, in which the two-electron SE diagrams contribute, namely, He-like ions. Typically, the slow convergence of partial-wave expansions becomes more pronounced as $Z$ decreases. For these reasons, the most suitable candidates to probe the nonperturbative (in $\alpha Z$) methods are middle-$Z$ systems. The most recent and accurate calculations of the two-electron SE contributions to energy levels of middle-$Z$ He-like ions were performed in Ref.~\cite{Yerokhin:2022:022815} and covered the the range $10 \leqslant Z\leqslant 40$. In the present work, we use the results of Ref.~\cite{Yerokhin:2022:022815} as the reference ones and perform the calculations for the $(1s1s)_0$, $(1s2p_{1/2})_0$, and $(1s2p_{3/2})_2$ states in He-like neon ($Z=10$), sulfur ($Z=16$), chromium ($Z=24$), and germanium ($Z=32$). As in Ref.~\cite{Yerokhin:2022:022815}, the calculations are carried out for the point-nucleus Coulomb potential, $V_{\rm C}(x)=-\alpha Z/x$. The Feynman gauge is used for the photon propagator. All the results are presented in terms of the dimensionless function $F(\alpha Z)$ defined by
\begin{align}
\label{eq:F2}
E = \alpha^2 ( \alpha Z )^3 F( \alpha Z ) \, mc^2 \, .
\end{align}

\begin{table*}[t]
\centering

\renewcommand{\arraystretch}{1.3}

\caption{\label{tab:irA_Ne} 
         Individual contributions to the sum of irreducible, $E_{\rm irr}$, and A-reducible, $E_{\rm red}^{\rm A}$, terms for the 
         $(1s1s)_0$, $(1s2p_{1/2})_0$, and $(1s2p_{3/2})_2$ states in He-like neon ($Z=10$),
         in terms of the function $F(\alpha Z)$ defined in Eq.~(\ref{eq:F2}).
         ``I'' denotes the standard potential-expansion approach shown in Fig.~\ref{fig:calc_1}, 
         ``II'' indicates the subtraction scheme in Fig.~\ref{fig:calc_4} proposed in Ref.~\cite{Sapirstein:2023:042804}, whereas 
         ``III'' stands for the scheme in Fig.~\ref{fig:calc_5}.
         }
         
\resizebox{\textwidth}{!}{%
\begin{tabular}{
                c
                S[table-format=-4.5(2),group-separator=,table-align-text-post=false]
                S[table-format=-2.6(1),group-separator=,table-align-text-post=false]
                S[table-format=-2.7(1),group-separator=,table-align-text-post=false]
                S[table-format=-3.5(2),group-separator=,table-align-text-post=false]
                S[table-format=-1.6(1),group-separator=,table-align-text-post=false]
                S[table-format=-1.7(1),group-separator=,table-align-text-post=false]
                S[table-format=-3.5(2),group-separator=,table-align-text-post=false]
                S[table-format=-1.6(1),group-separator=,table-align-text-post=false]
                S[table-format=-1.6,group-separator=,table-align-text-post=false]
               }
               
\hline
\hline

                                                               &
  \multicolumn{3}{c}{\rule{0pt}{2.6ex}$~~~~~(1s1s)_0$}            &
  \multicolumn{3}{c}{                  $~~~(1s2p_{1/2})_0$}      &
  \multicolumn{3}{c}{                  $~~~~(1s2p_{3/2})_2$}     \\ 
  
                                                        &
  \multicolumn{1}{c}{I}       &
  \multicolumn{1}{c}{II}      &
  \multicolumn{1}{c}{III}     &  
  \multicolumn{1}{c}{I}       &
  \multicolumn{1}{c}{II}      &
  \multicolumn{1}{c}{III}     &  
  \multicolumn{1}{c}{I}       &
  \multicolumn{1}{c}{II}      &
  \multicolumn{1}{c}{III}     \\  
        
\hline   
                       
 Free        &    26.066354 &    26.066354 &    26.066354 &    11.037001 &    11.037001 &    11.037001 &    10.882193 &    10.882193 &    10.882193   \rule{0pt}{3.2ex}   \\ 
 Subtraction &              &   -16.894640 &   -13.802631 &              &    -6.447367 &    -5.620968 &              &    -6.416416 &    -5.574568   \\ 
 $|\kappa|=1$        &   -27.382533 &   -11.063974 &   -14.147369 &   -10.722851 &    -4.668301 &    -5.495020 &    -2.240635 &    -0.895486 &    -1.097068   \\ 
  2        &    -0.446027 &    -0.034606 &    -0.043049 &    -0.262862 &    -0.027852 &    -0.027515 &    -8.454708 &    -3.684125 &    -4.319842   \\ 
  3        &    -0.104113 &    -0.004100 &    -0.004538 &    -0.074401 &    -0.003449 &    -0.003532 &    -0.194388 &    -0.012349 &    -0.016173   \\ 
  4        &    -0.037055 &    -0.000787 &    -0.000785 &    -0.033382 &    -0.001039 &    -0.001042 &    -0.054056 &    -0.001680 &    -0.002112   \\ 
  5        &    -0.015989 &    -0.000168 &    -0.000129 &    -0.018217 &    -0.000438 &    -0.000430 &    -0.024509 &    -0.000469 &    -0.000603   \\ 
  6        &    -0.007612 &    -0.000020 &     0.000017 &    -0.011072 &    -0.000220 &    -0.000211 &    -0.013581 &    -0.000179 &    -0.000240   \\ 
  7        &    -0.003807 &     0.000018 &     0.000048 &    -0.007208 &    -0.000124 &    -0.000116 &    -0.008380 &    -0.000080 &    -0.000113   \\ 
  8        &    -0.001929 &     0.000025 &     0.000049 &    -0.004923 &    -0.000075 &    -0.000068 &    -0.005531 &    -0.000039 &    -0.000059   \\ 
  9        &    -0.000951 &     0.000024 &     0.000043 &    -0.003483 &    -0.000048 &    -0.000043 &    -0.003823 &    -0.000020 &    -0.000033   \\ 
 10        &    -0.000425 &     0.000020 &     0.000036 &    -0.002532 &    -0.000032 &    -0.000028 &    -0.002734 &    -0.000010 &    -0.000019   \\ 
 11        &    -0.000137 &     0.000017 &     0.000030 &    -0.001881 &    -0.000022 &    -0.000018 &    -0.002006 &    -0.000005 &    -0.000011   \\ 
 12        &     0.000019 &     0.000014 &     0.000024 &    -0.001421 &    -0.000016 &    -0.000013 &    -0.001501 &    -0.000002 &    -0.000007   \\ 
 13        &     0.000103 &     0.000011 &     0.000020 &    -0.001089 &    -0.000011 &    -0.000009 &    -0.001142 &    -0.000001 &    -0.000004   \\ 
 14        &     0.000144 &     0.000009 &     0.000016 &    -0.000844 &    -0.000008 &    -0.000006 &    -0.000880 &    -0.000000 &    -0.000002   \\ 
 15        &     0.000161 &     0.000008 &     0.000014 &    -0.000660 &    -0.000006 &    -0.000004 &    -0.000686 &     0.000000 &    -0.000001   \\ 
 $\sum_{|\kappa|=16}^{25}$  &     0.001228 &     0.000033 &     0.000064 &    -0.002319 &    -0.000020 &    -0.000011 &    -0.002388 &     0.000008 &     0.000002   \\ 
 $\sum_{|\kappa|>25}{\rm[extr.]}$  &  0.00100(27) &  0.000016(1) &     0.000039 &  0.00006(55) & -0.000001(1) &  0.000005(1) &  0.00004(54) &  0.000010(2) &     0.000010   \\ 
 $E_{\rm irA}$  & -1.93157(27) & -1.931745(1) &    -1.931745 & -0.11209(55) & -0.112028(1) & -0.112028(1) & -0.12871(54) & -0.128651(2) &    -0.128652   \\ 
 Ref.~\cite{Yerokhin:private}  &   &   &   &   &    -0.112028 &   &   &  -0.128652(1) &     \\ 

\hline
\hline

\end{tabular}%
}

\end{table*}

\begin{table*}[t]
\centering

\renewcommand{\arraystretch}{1.3}

\caption{\label{tab:SC1_xp_Ne} 
         Subtractions for the B-reducible, $E_{\rm red}^{\rm B}$, and vertex, $E_{\rm vert}$, terms in He-like neon ($Z=10$),
         in terms of the function $F(\alpha Z)$ defined in Eq.~(\ref{eq:F2}).
         Comparison of the coordinate-space (X) and momentum-space (P) calculations.
         }
         
\resizebox{\textwidth}{!}{%
\begin{tabular}{
                c
                S[table-format=5.7(2),group-separator=,table-align-text-post=false]
                S[table-format=-3.7(2),group-separator=,table-align-text-post=false]
                S[table-format=5.6(2),group-separator=,table-align-text-post=false]
                S[table-format=-3.6(2),group-separator=,table-align-text-post=false]
                S[table-format=5.6(2),group-separator=,table-align-text-post=false]
                S[table-format=-3.6(2),group-separator=,table-align-text-post=false]
               }
               
\hline
\hline

                                                              &
  \multicolumn{2}{c}{\rule{0pt}{3.2ex}$(1s1s)_0$}           &
  \multicolumn{2}{c}{                  $(1s2p_{1/2})_0$}     &
  \multicolumn{2}{c}{                  $(1s2p_{3/2})_2$}     \\ 
  
                                                        & 
  \multicolumn{1}{c}{$E_{\rm red}^{\rm B}$}       &
  \multicolumn{1}{c}{$E_{\rm vert}$}       &
  \multicolumn{1}{c}{$E_{\rm red}^{\rm B}$}       &
  \multicolumn{1}{c}{$E_{\rm vert}$}       &
  \multicolumn{1}{c}{$E_{\rm red}^{\rm B}$}       &
  \multicolumn{1}{c}{$E_{\rm vert}$}       \\[0.5mm]        
        
\hline   
                       
 $|\kappa|=1$        &   66.2764394 &  -66.0951856 &   24.1381053 &  -24.1663346 &   12.2942539 &  -13.1840049  \rule{0pt}{3.2ex}   \\ 
  2        &    2.8338987 &   -2.7247985 &    1.1358243 &   -1.0746620 &   12.4466358 &  -11.5650413   \\ 
  3        &    0.8124463 &   -0.7635509 &    0.3382531 &   -0.3222053 &    0.6752695 &   -0.6356262   \\ 
  4        &    0.3663848 &   -0.3378507 &    0.1596131 &   -0.1519373 &    0.2200592 &   -0.2085721   \\ 
  5        &    0.2010034 &   -0.1822806 &    0.0914882 &   -0.0868552 &    0.1108712 &   -0.1051131   \\ 
  6        &    0.1230834 &   -0.1099815 &    0.0584104 &   -0.0552524 &    0.0665551 &   -0.0629485   \\ 
  7        &    0.0809627 &   -0.0713990 &    0.0399785 &   -0.0376627 &    0.0439850 &   -0.0414565   \\ 
  8        &    0.0560434 &   -0.0488455 &    0.0287400 &   -0.0269582 &    0.0309280 &   -0.0290337   \\ 
  9        &    0.0403193 &   -0.0347725 &    0.0214344 &   -0.0200166 &    0.0227223 &   -0.0212405   \\ 
 10        &    0.0299022 &   -0.0255456 &    0.0164511 &   -0.0152945 &    0.0172534 &   -0.0160585   \\ 
 $\sum_{|\kappa|=11}^{20}$  &    0.1014834 &   -0.0844362 &    0.0627488 &   -0.0573333 &    0.0644667 &   -0.0589901   \\ 
 $\sum_{|\kappa|=21}^{30}$  &    0.0187646 &   -0.0150319 &    0.0145483 &   -0.0127979 &    0.0146678 &   -0.0129256   \\ 
 $\sum_{|\kappa|=31}^{40}$  &    0.0061474 &   -0.0048819 &    0.0054127 &   -0.0046375 &    0.0054314 &   -0.0046625   \\ 
 $\sum_{|\kappa|=41}^{50}$  &    0.0026757 &   -0.0021282 &    0.0025253 &   -0.0021250 &    0.0025290 &   -0.0021326   \\ 
 $\sum_{|\kappa|>50}{\rm[extr.]}$  & 0.004235(53) & -0.003421(30) &  0.00425(24) & -0.00348(19) &  0.00424(24) & -0.00349(19)   \\ 
 X-space  & 70.953789(53) & -70.504109(30) & 26.11778(24) & -26.03755(19) & 26.01987(24) & -25.95129(19)   \\ 
 P-space  &   70.9538063 &  -70.5041268 &   26.1176417 &  -26.0374674 &   26.0197341 &  -25.9512077   \\ 

\hline
\hline

\end{tabular}%
}

\end{table*}

\begin{table*}[t]
\centering

\renewcommand{\arraystretch}{1.3}

\caption{\label{tab:SC1_xp_U_} 
         The same as in Table~\ref{tab:SC1_xp_Ne} for He-like uranium ($Z=92$).
         }
         
\begin{tabular}{
                c
                S[table-format=4.8,group-separator=,table-align-text-post=false]
                S[table-format=-2.8,group-separator=,table-align-text-post=false]
                S[table-format=4.8(1),group-separator=,table-align-text-post=false]
                S[table-format=-2.8(1),group-separator=,table-align-text-post=false]
                S[table-format=4.8(1),group-separator=,table-align-text-post=false]
                S[table-format=-2.8(1),group-separator=,table-align-text-post=false]
               }
               
\hline
\hline

                                                              &
  \multicolumn{2}{c}{\rule{0pt}{3.2ex}$(1s1s)_0$}           &
  \multicolumn{2}{c}{                  $(1s2p_{1/2})_0$}     &
  \multicolumn{2}{c}{                  $(1s2p_{3/2})_2$}     \\ 
  
                                                        & 
  \multicolumn{1}{c}{$E_{\rm red}^{\rm B}$}       &
  \multicolumn{1}{c}{$E_{\rm vert}$}       &
  \multicolumn{1}{c}{$E_{\rm red}^{\rm B}$}       &
  \multicolumn{1}{c}{$E_{\rm vert}$}       &
  \multicolumn{1}{c}{$E_{\rm red}^{\rm B}$}       &
  \multicolumn{1}{c}{$E_{\rm vert}$}       \\[0.5mm]         
        
\hline   
                       
 $|\kappa|=1$        &    0.4863571 &   -0.5007927 &    0.2059600 &   -0.1940478 &    0.0785919 &   -0.0779925  \rule{0pt}{3.2ex}   \\ 
  2        &    0.0121825 &   -0.0117173 &    0.0173520 &   -0.0084045 &    0.0755508 &   -0.0704169   \\ 
  3        &    0.0103269 &   -0.0091830 &    0.0086529 &   -0.0058546 &    0.0121050 &   -0.0108155   \\ 
  4        &    0.0051811 &   -0.0045102 &    0.0042482 &   -0.0028544 &    0.0048203 &   -0.0041160   \\ 
  5        &    0.0028344 &   -0.0024423 &    0.0023265 &   -0.0015356 &    0.0023598 &   -0.0019708   \\ 
  6        &    0.0016945 &   -0.0014524 &    0.0013909 &   -0.0009043 &    0.0013196 &   -0.0010880   \\ 
  7        &    0.0010868 &   -0.0009288 &    0.0008894 &   -0.0005717 &    0.0008074 &   -0.0006606   \\ 
  8        &    0.0007363 &   -0.0006282 &    0.0005995 &   -0.0003821 &    0.0005272 &   -0.0004295   \\ 
  9        &    0.0005210 &   -0.0004441 &    0.0004215 &   -0.0002670 &    0.0003619 &   -0.0002941   \\ 
 10        &    0.0003817 &   -0.0003252 &    0.0003067 &   -0.0001934 &    0.0002584 &   -0.0002098   \\ 
 $\sum_{|\kappa|=11}^{20}$  &    0.0012845 &   -0.0010942 &    0.0010080 &   -0.0006315 &    0.0008130 &   -0.0006622   \\ 
 $\sum_{|\kappa|=21}^{30}$  &    0.0002528 &   -0.0002156 &    0.0001907 &   -0.0001193 &    0.0001453 &   -0.0001199   \\ 
 $\sum_{|\kappa|=31}^{40}$  &    0.0000904 &   -0.0000771 &    0.0000671 &   -0.0000421 &    0.0000502 &   -0.0000417   \\ 
 $\sum_{|\kappa|=41}^{50}$  &    0.0000423 &   -0.0000361 &    0.0000312 &   -0.0000196 &    0.0000231 &   -0.0000193   \\ 
 $\sum_{|\kappa|>50}{\rm[extr.]}$  &    0.0000763 &   -0.0000652 & 0.0000558(1) & -0.0000352(2) & 0.0000410(1) & -0.0000344(1)   \\ 
 X-space  &    0.5230486 &   -0.5339123 & 0.2435003(1) & -0.2158631(2) & 0.1777750(1) & -0.1688713(1)   \\ 
 P-space  &    0.5230486 &   -0.5339124 &    0.2435004 &   -0.2158632 &    0.1777750 &   -0.1688713   \\ 

\hline
\hline

\end{tabular}%

\end{table*}

\begin{table*}[t]
\centering

\renewcommand{\arraystretch}{1.3}

\caption{\label{tab:scrse_1s1s_0} 
         Individual contributions to the two-electron self-energy correction for the $(1s1s)_0$ state in He-like ions,
         in terms of the function $F(\alpha Z)$ defined in Eq.~(\ref{eq:F2}).
         $E_{\rm irA}$ stands for the sum of irreducible and A-reducible terms. All the other rows are related with the B-reducible and vertex terms. ``I'' denotes the standard potential-expansion approach, whereas ``II'' indicates the new subtraction scheme. 
         }
         
\resizebox{\textwidth}{!}{%
\begin{tabular}{
                c
                S[table-format=-3.6(2),group-separator=,table-align-text-post=false]
                S[table-format=-2.6(2),group-separator=,table-align-text-post=false]
                S[table-format=-3.6(2),group-separator=,table-align-text-post=false]
                S[table-format=-2.6(2),group-separator=,table-align-text-post=false]
                S[table-format=-3.6(2),group-separator=,table-align-text-post=false]
                S[table-format=-2.6(2),group-separator=,table-align-text-post=false]
                S[table-format=-3.6(2),group-separator=,table-align-text-post=false]
                S[table-format=-2.6(2),group-separator=,table-align-text-post=false]
               }
               
\hline
\hline

                                                    &
  \multicolumn{2}{c}{\rule{0pt}{2.6ex}$Z=10$}     &
  \multicolumn{2}{c}{                  $Z=16$}     &
  \multicolumn{2}{c}{                  $Z=24$}     &
  \multicolumn{2}{c}{                  $Z=32$}     \\ 
  
                                                        & 
  \multicolumn{1}{c}{I}       &
  \multicolumn{1}{c}{II}      & 
  \multicolumn{1}{c}{I}       &
  \multicolumn{1}{c}{II}      & 
  \multicolumn{1}{c}{I}       &
  \multicolumn{1}{c}{II}      &  
  \multicolumn{1}{c}{I}       &
  \multicolumn{1}{c}{II}      \\ 
        
\hline   
                       
 Free        &    10.250914 &    10.250914 &     3.674285 &     3.674285 &     1.431102 &     1.431102 &     0.690929 &     0.690929   \rule{0pt}{3.2ex}    \\ 
 Subtraction &              &     0.449680 &              &     0.287918 &              &     0.175509 &              &     0.112889   \\ 
 $|\kappa|=1$        &   -10.983633 &   -11.164887 &    -4.228012 &    -4.350231 &    -1.856526 &    -1.934592 &    -1.044016 &    -1.095262   \\ 
  2        &     0.094565 &    -0.014535 &     0.065496 &    -0.006481 &     0.042436 &    -0.002292 &     0.028339 &    -0.000938   \\ 
  3        &     0.048155 &    -0.000740 &     0.033445 &     0.000407 &     0.021595 &     0.000935 &     0.014501 &     0.000989   \\ 
  4        &     0.028598 &     0.000064 &     0.018758 &     0.000405 &     0.011421 &     0.000515 &     0.007343 &     0.000478   \\ 
  5        &     0.018860 &     0.000137 &     0.011650 &     0.000264 &     0.006682 &     0.000281 &     0.004113 &     0.000240   \\ 
  6        &     0.013220 &     0.000118 &     0.007719 &     0.000170 &     0.004193 &     0.000162 &     0.002484 &     0.000130   \\ 
  7        &     0.009655 &     0.000091 &     0.005352 &     0.000112 &     0.002770 &     0.000098 &     0.001588 &     0.000075   \\ 
  8        &     0.007266 &     0.000068 &     0.003842 &     0.000076 &     0.001905 &     0.000062 &     0.001063 &     0.000045   \\ 
  9        &     0.005598 &     0.000052 &     0.002834 &     0.000053 &     0.001354 &     0.000041 &     0.000738 &     0.000028   \\ 
 10        &     0.004396 &     0.000039 &     0.002139 &     0.000037 &     0.000988 &     0.000028 &     0.000528 &     0.000018   \\ 
 11        &     0.003507 &     0.000030 &     0.001645 &     0.000027 &     0.000738 &     0.000019 &     0.000389 &     0.000012   \\ 
 12        &     0.002835 &     0.000024 &     0.001287 &     0.000020 &     0.000563 &     0.000014 &     0.000292 &     0.000008   \\ 
 $\sum_{|\kappa|=13}^{18}$  &     0.009287 &     0.000069 &     0.003888 &     0.000053 &     0.001611 &     0.000032 &     0.000818 &     0.000018   \\ 
 $\sum_{|\kappa|>18}{\rm[extr.]}$  &  0.00862(70) & 0.000045(12) &  0.00297(39) &  0.000024(8) &  0.00111(12) &  0.000010(3) & 0.000555(35) &  0.000002(1)   \\ 
 $E_{\rm vr}$  & -0.47816(70) & -0.478833(12) & -0.39270(39) & -0.392861(8) & -0.32805(12) & -0.328076(3) & -0.290335(35) & -0.290337(1)   \\ 
 $E_{\rm irA}$  &    -1.931745 &    -1.931745 &    -1.528847 &    -1.528847 &    -1.230200 &    -1.230200 &    -1.054395 &    -1.054395   \\ 
 Total  & -2.40990(70) & -2.410578(12) & -1.92155(39) & -1.921708(8) & -1.55825(12) & -1.558275(3) & -1.344730(35) & -1.344732(1)   \\ 
 Ref.~\cite{Yerokhin:2022:022815}  &              & -2.41058(36) &              &  -1.92171(9) &              &  -1.55827(6) &              &  -1.34472(3)   \\ 

\hline
\hline

\end{tabular}%
}

\end{table*}

\begin{table*}[t]
\centering

\renewcommand{\arraystretch}{1.3}

\caption{\label{tab:scrse_1s2p1_0} 
         Individual contributions to the two-electron self-energy correction for the $(1s2p_{1/2})_0$ state in He-like ions,
         in terms of the function $F(\alpha Z)$ defined in Eq.~(\ref{eq:F2}).
         The notations are the same as in Table~\ref{tab:scrse_1s1s_0}.
         }
         
\resizebox{\textwidth}{!}{%
\begin{tabular}{
                c
                S[table-format=-3.6(2),group-separator=,table-align-text-post=false]
                S[table-format=-2.6(2),group-separator=,table-align-text-post=false]
                S[table-format=-3.6(2),group-separator=,table-align-text-post=false]
                S[table-format=-2.6(2),group-separator=,table-align-text-post=false]
                S[table-format=-3.6(2),group-separator=,table-align-text-post=false]
                S[table-format=-2.6(2),group-separator=,table-align-text-post=false]
                S[table-format=-3.6(2),group-separator=,table-align-text-post=false]
                S[table-format=-2.6(2),group-separator=,table-align-text-post=false]
               }
               
\hline
\hline

                                                    &
  \multicolumn{2}{c}{\rule{0pt}{2.6ex}$Z=10$}     &
  \multicolumn{2}{c}{                  $Z=16$}     &
  \multicolumn{2}{c}{                  $Z=24$}     &
  \multicolumn{2}{c}{                  $Z=32$}     \\ 
  
                                                        & 
  \multicolumn{1}{c}{I}       &
  \multicolumn{1}{c}{II}      & 
  \multicolumn{1}{c}{I}       &
  \multicolumn{1}{c}{II}      & 
  \multicolumn{1}{c}{I}       &
  \multicolumn{1}{c}{II}      &  
  \multicolumn{1}{c}{I}       &
  \multicolumn{1}{c}{II}      \\ 
        
\hline   
                       
 Free        &     2.788828 &     2.788828 &     1.094813 &     1.094813 &     0.495255 &     0.495255 &     0.285677 &     0.285677   \rule{0pt}{3.2ex}    \\ 
 Subtraction &              &     0.080174 &              &     0.063992 &              &     0.052388 &              &     0.045659   \\ 
 $|\kappa|=1$        &    -2.895668 &    -2.867439 &    -1.162282 &    -1.157730 &    -0.540583 &    -0.547555 &    -0.320835 &    -0.332652   \\ 
  2        &     0.069329 &     0.008167 &     0.045089 &     0.007713 &     0.031639 &     0.007338 &     0.025154 &     0.007137   \\ 
  3        &     0.016377 &     0.000329 &     0.011240 &     0.000520 &     0.008157 &     0.000701 &     0.006574 &     0.000840   \\ 
  4        &     0.007763 &     0.000088 &     0.005501 &     0.000172 &     0.004056 &     0.000253 &     0.003274 &     0.000313   \\ 
  5        &     0.004678 &     0.000045 &     0.003350 &     0.000090 &     0.002462 &     0.000133 &     0.001965 &     0.000161   \\ 
  6        &     0.003187 &     0.000029 &     0.002278 &     0.000057 &     0.001652 &     0.000081 &     0.001297 &     0.000095   \\ 
  7        &     0.002337 &     0.000021 &     0.001655 &     0.000039 &     0.001179 &     0.000053 &     0.000907 &     0.000061   \\ 
  8        &     0.001798 &     0.000016 &     0.001257 &     0.000028 &     0.000876 &     0.000037 &     0.000661 &     0.000041   \\ 
  9        &     0.001431 &     0.000013 &     0.000984 &     0.000021 &     0.000672 &     0.000027 &     0.000496 &     0.000028   \\ 
 10        &     0.001167 &     0.000010 &     0.000789 &     0.000016 &     0.000527 &     0.000020 &     0.000382 &     0.000021   \\ 
 11        &     0.000970 &     0.000009 &     0.000644 &     0.000013 &     0.000421 &     0.000015 &     0.000299 &     0.000015   \\ 
 12        &     0.000819 &     0.000007 &     0.000534 &     0.000010 &     0.000341 &     0.000012 &     0.000238 &     0.000011   \\ 
 $\sum_{|\kappa|=13}^{18}$  &     0.003060 &     0.000025 &     0.001886 &     0.000033 &     0.001133 &     0.000034 &     0.000754 &     0.000031   \\ 
 $\sum_{|\kappa|>18}{\rm[extr.]}$  &  0.00417(71) &  0.000028(3) &  0.00219(13) &  0.000028(6) & 0.001108(75) &  0.000022(6) & 0.000652(63) &  0.000017(5)   \\ 
 $E_{\rm vr}$  &  0.01024(71) &  0.010352(3) &  0.00993(13) &  0.009815(6) & 0.008894(75) &  0.008814(6) & 0.007496(63) &  0.007454(5)   \\ 
 $E_{\rm irA}$  & -0.112028(1) & -0.112028(1) &    -0.088136 &    -0.088136 & -0.072143(1) & -0.072143(1) & -0.064514(1) & -0.064514(1)   \\ 
 Total  & -0.10179(71) & -0.101676(3) & -0.07821(13) & -0.078321(6) & -0.063249(75) & -0.063329(6) & -0.057018(63) & -0.057060(5)   \\ 
 Ref.~\cite{Yerokhin:2022:022815}  &              &  -0.10163(5) &              &  -0.07830(4) &              &  -0.06333(2) &              &  -0.05705(1)   \\ 

\hline
\hline

\end{tabular}%
}

\end{table*}

\begin{table*}[t]
\centering

\renewcommand{\arraystretch}{1.3}

\caption{\label{tab:scrse_1s2p3_2} 
         Individual contributions to the two-electron self-energy correction for the $(1s2p_{3/2})_2$ state in He-like ions,
         in terms of the function $F(\alpha Z)$ defined in Eq.~(\ref{eq:F2}).
         The notations are the same as in Table~\ref{tab:scrse_1s1s_0}.
         }
         
\resizebox{\textwidth}{!}{%
\begin{tabular}{
                c
                S[table-format=-3.6(2),group-separator=,table-align-text-post=false]
                S[table-format=-2.6(2),group-separator=,table-align-text-post=false]
                S[table-format=-3.6(2),group-separator=,table-align-text-post=false]
                S[table-format=-2.6(2),group-separator=,table-align-text-post=false]
                S[table-format=-3.6(2),group-separator=,table-align-text-post=false]
                S[table-format=-2.6(2),group-separator=,table-align-text-post=false]
                S[table-format=-3.6(2),group-separator=,table-align-text-post=false]
                S[table-format=-2.6(2),group-separator=,table-align-text-post=false]
               }
               
\hline
\hline

                                                    &
  \multicolumn{2}{c}{\rule{0pt}{2.6ex}$Z=10$}     &
  \multicolumn{2}{c}{                  $Z=16$}     &
  \multicolumn{2}{c}{                  $Z=24$}     &
  \multicolumn{2}{c}{                  $Z=32$}     \\ 
  
                                                        & 
  \multicolumn{1}{c}{I}       &
  \multicolumn{1}{c}{II}      & 
  \multicolumn{1}{c}{I}       &
  \multicolumn{1}{c}{II}      & 
  \multicolumn{1}{c}{I}       &
  \multicolumn{1}{c}{II}      &  
  \multicolumn{1}{c}{I}       &
  \multicolumn{1}{c}{II}      \\ 
        
\hline   
                       
 Free        &     2.724270 &     2.724270 &     1.033948 &     1.033948 &     0.439756 &     0.439756 &     0.235699 &     0.235699   \rule{0pt}{3.2ex}    \\ 
 Subtraction &              &     0.068526 &              &     0.050658 &              &     0.037062 &              &     0.028743   \\ 
 $|\kappa|=1$        &   -16.550482 &   -15.660731 &    -6.734048 &    -6.417569 &    -3.049640 &    -2.927424 &    -1.709396 &    -1.650886   \\ 
  2        &    13.731826 &    12.850231 &     5.635484 &     5.314444 &     2.562252 &     2.431217 &     1.434352 &     1.366290   \\ 
  3        &     0.039937 &     0.000293 &     0.022978 &     0.000503 &     0.013757 &     0.000601 &     0.009297 &     0.000623   \\ 
  4        &     0.011577 &     0.000090 &     0.007677 &     0.000197 &     0.005212 &     0.000272 &     0.003851 &     0.000307   \\ 
  5        &     0.005801 &     0.000043 &     0.003994 &     0.000090 &     0.002771 &     0.000123 &     0.002066 &     0.000139   \\ 
  6        &     0.003635 &     0.000028 &     0.002523 &     0.000052 &     0.001742 &     0.000069 &     0.001286 &     0.000076   \\ 
  7        &     0.002548 &     0.000020 &     0.001760 &     0.000034 &     0.001198 &     0.000043 &     0.000870 &     0.000046   \\ 
  8        &     0.001909 &     0.000015 &     0.001304 &     0.000024 &     0.000871 &     0.000029 &     0.000621 &     0.000030   \\ 
  9        &     0.001493 &     0.000012 &     0.001005 &     0.000017 &     0.000657 &     0.000020 &     0.000460 &     0.000021   \\ 
 10        &     0.001204 &     0.000009 &     0.000797 &     0.000013 &     0.000510 &     0.000015 &     0.000350 &     0.000015   \\ 
 11        &     0.000993 &     0.000007 &     0.000645 &     0.000010 &     0.000404 &     0.000011 &     0.000272 &     0.000011   \\ 
 12        &     0.000833 &     0.000006 &     0.000532 &     0.000008 &     0.000326 &     0.000008 &     0.000216 &     0.000008   \\ 
 $\sum_{|\kappa|=13}^{18}$  &     0.003078 &     0.000020 &     0.001858 &     0.000024 &     0.001072 &     0.000024 &     0.000676 &     0.000021   \\ 
 $\sum_{|\kappa|>18}{\rm[extr.]}$  &  0.00413(70) &  0.000020(2) &  0.00212(13) &  0.000019(4) & 0.001032(72) &  0.000015(3) & 0.000573(59) &  0.000012(3)   \\ 
 $E_{\rm vr}$  & -0.01725(70) & -0.017141(2) & -0.01742(13) & -0.017528(4) & -0.018080(72) & -0.018159(3) & -0.018806(59) & -0.018846(3)   \\ 
 $E_{\rm irA}$  &    -0.128652 &    -0.128652 & -0.102186(1) & -0.102186(1) & -0.082198(1) & -0.082198(1) & -0.070101(1) & -0.070101(1)   \\ 
 Total  & -0.14590(70) & -0.145792(2) & -0.11960(13) & -0.119714(4) & -0.100278(72) & -0.100357(3) & -0.088907(59) & -0.088948(3)   \\ 
 Ref.~\cite{Yerokhin:2022:022815}  &              & -0.14574(12) &              &  -0.11969(7) &              &  -0.10035(5) &              &  -0.08894(4)   \\ 

\hline
\hline

\end{tabular}%
}

\end{table*}

Let us start with the contribution~$E_{\rm irA}$. As noted above, this contribution can be evaluated using the approaches developed to treat the first-order SE correction~(\ref{eq:SE}). In Ref.~\cite{Yerokhin:2022:022815}, e.g., it was calculated by means of the method proposed in Ref.~\cite{Yerokhin:2005:042502}. In Table~\ref{tab:irA_Ne}, we present our results for the case of $Z=10$, which is the most difficult in terms of convergence. For all the states, we compare three calculation schemes: column~``I'' shows the standard PE approach given in Fig.~\ref{fig:calc_1}, while columns~``II'' and ``III'' present the results obtained within the convergence-acceleration methods depicted in Figs.~\ref{fig:calc_4} and \ref{fig:calc_5}, respectively. The line labeled ``Free'' is common to all the schemes and shows the sums of the zero- and one-potential contributions evaluated in momentum space. ``Subtraction'' stands for the P-space terms in Figs.~\ref{fig:calc_4} and \ref{fig:calc_5}, they are absent in the standard approach. The subsequent rows show the individual partial-wave-expansion contributions for different values of $|\kappa|$. For all the schemes, we truncate the calculations at $|\kappa_{\rm max}|=25$. The line labeled ``$\sum_{|\kappa|>25}$[extr.]'' gives the estimates for the remainders of the partial-wave series. The total values are presented in the row~$E_{\rm irA}$. Here and below, the numbers in parentheses are the uncertainties in the last digits. If no uncertainties are given, numerical values are assumed to be accurate to all digits specified. We note that in all the cases the uncertainties of the calculations are solely due to the extrapolation procedure.

As can be seen from Table~\ref{tab:irA_Ne}, the standard PE approach to the contribution~$E_{\rm irA}$ indeed suffers from the slow convergence of the partial-wave expansion. Accurate and reliable extrapolation is difficult in this case. The convergence-acceleration methods correct the situation significantly. We stress that the two considered methods are in excellent agreement with each other despite the fact that their ``Subtraction'' terms have completely different values. We also note that from the cases considered in Table~\ref{tab:irA_Ne} it is not possible to conclude that one method is superior to the other. For the excited $(1s2p_{1/2})_0$ and $(1s2p_{3/2})_2$ states, the comparison with the results from Refs.~\cite{Yerokhin:2022:022815, Yerokhin:private}, obtained within the approach of Ref.~\cite{Yerokhin:2005:042502}, is also given. Excellent agreement is found, see also the related discussion below.

Let us now pass to the contribution~$E_{\rm vr}$ which is of primary interest. For the vertex contribution~$E_{\rm vert}$ in Fig.~\ref{fig:calc_6} and for the B-reducible contribution~$E_{\rm red}^{\rm B}$ in the similar scheme, the subtraction terms appear in two forms, X- and P-space ones. As noted above, the subtractions in coordinate space are calculated within the partial-wave expansion using the same techniques as employed for the X-space terms in Figs.~\ref{fig:calc_2} and \ref{fig:calc_3}. The subtractions in momentum space are evaluated in the closed form of multidimensional integrals, see Appendices~\ref{sec:app:2} and~\ref{sec:app:3}. To cross-check our methods, in Tables~\ref{tab:SC1_xp_Ne} and \ref{tab:SC1_xp_U_} we compare the X- and P-space values of the subtractions for He-like neon ($Z=10$) and uranium ($Z=92$), respectively. This choice of $Z$ serves to demonstrate how the rate of partial-wave-expansion convergence changes along the isoelectronic sequence: the lower $Z$, the worse the convergence. Tables~\ref{tab:SC1_xp_Ne} and \ref{tab:SC1_xp_U_} are organized as follows. At first, the individual partial-wave-expansion contributions obtained within the X-space calculations for different values of $\kappa$ are shown. These calculations are truncated at $|\kappa_{\rm max}|=50$. The lines labeled ``$\sum_{|\kappa|>50}$[extr.]'' provide the partial-wave-series remainders obtained by extrapolation. The total X-space values are in the penultimate rows. The P-space values of the subtractions are shown in the last lines. 

The data in Table~\ref{tab:SC1_xp_Ne} once again confirm that the PE calculations for low-$Z$ systems are a challenging problem. Even consideration of all partial waves with $|\kappa|\leqslant 50$, supplemented by the extrapolation of the obtained results, is significantly inferior in accuracy to the corresponding P-space evaluation. Nevertheless, both ways to calculate the subtraction terms are in good agreement. As can be seen from Table~\ref{tab:SC1_xp_U_}, the partial-wave-convergence situation improves considerably for high-$Z$ ions. The perfect agreement between the X- and P-space values is found in this case. The coincidence of the results obtained by means of completely different methods is a good test of the used numerical procedures.

In Tables~\ref{tab:scrse_1s1s_0}, \ref{tab:scrse_1s2p1_0}, and \ref{tab:scrse_1s2p3_2}, we present the details of the calculations of the contribution~$E_{\rm vr}$ for the $(1s1s)_0$, $(1s2p_{1/2})_0$, and $(1s2p_{3/2})_2$ states, respectively. For all the considered states and values of $Z$, we compare two calculation schemes: column ``I'' stands for the standard PE approach shown in Figs.~\ref{fig:calc_2} and \ref{fig:calc_3}, while column ``II'' presents the performance of our convergence-acceleration method. The row labeled ``Free'' shows the sums of the P-space terms in Figs.~\ref{fig:calc_2} and \ref{fig:calc_3} resulting from the renormalization procedure; these contributions are common to both schemes. For the convergence-acceleration approach, the line ``Subtraction'' gives the corresponding contributions evaluated in momentum space. The next lines present the individual contributions of the partial-wave expansions. In this case, the calculations are terminated at $|\kappa_{\rm max}|=18$. The row labeled ``$\sum_{|\kappa|>18}$[extr.]'' contains the partial-wave-series tails obtained by extrapolation. The total results for the sums of B-reducible and vertex terms are shown in the line $E_{\rm vr}$. From Tables~\ref{tab:scrse_1s1s_0}-\ref{tab:scrse_1s2p3_2}, one can see a drastic improvement in accuracy due to the application of the proposed convergence-acceleration method. 

To obtain the total value of the two-electron SE contribution, one has to add the contributions~$E_{\rm irA}$ and $E_{\rm vr}$. For this purpose, the contribution~$E_{\rm irA}$ has been calculated for $Z=16$, 24, and 32 according to the scheme shown in Fig.~\ref{fig:calc_5}, as realized for neon in Table~\ref{tab:irA_Ne}. The corresponding values are presented in Tables~\ref{tab:scrse_1s1s_0}-\ref{tab:scrse_1s2p3_2} in the row labeled~$E_{\rm irA}$; the same value of $E_{\rm irA}$ is added for both schemes for calculating~$E_{\rm vr}$. The resulting two-electron SE contributions are shown in the line ``Total''. The uncertainties of $E_{\rm irA}$ and $E_{\rm vr}$ are summed quadratically. However, in all the cases, the total uncertainties are determined by the accuracy with which the contribution~$E_{\rm vr}$ is calculated. 

In Tables~\ref{tab:scrse_1s1s_0}-\ref{tab:scrse_1s2p3_2}, we compare our results with those obtained in Ref.~\cite{Yerokhin:2022:022815}. In that work, the general scheme for treating the two-electron SE diagrams also follows the methods outlined in Refs.~\cite{Yerokhin:1999:800, Yerokhin:1999:3522}. For the irreducible and reducible contributions, the modification proposed in Ref.~\cite{Yerokhin:2005:042502} was used. The vertex contribution, which is the main source of the numerical uncertainty, was calculated employing the technique described in detail in Ref.~\cite{Yerokhin:2020:800}. No convergence-acceleration methods were applied in this case. To overcome the slow convergence of the partial-wave expansion for the vertex contribution, the calculations in Ref.~\cite{Yerokhin:2022:022815} were extended up to $|\kappa_{\rm max}|=50$. As one can see from Tables~\ref{tab:scrse_1s1s_0}-\ref{tab:scrse_1s2p3_2}, our results obtained within the convergence-acceleration approach are in excellent agreement with the ones from Ref.~\cite{Yerokhin:2022:022815} but have higher accuracy. 

Finally, we should note that extending the calculations of the contribution~$E_{\rm vr}$ within the convergence-acceleration method up to $|\kappa_{\rm max}|=18$ is, in some sense, excessive. For instance, if the calculations for $Z=10$ were truncated at $|\kappa_{\rm max}|=12$, we would obtain $-2.410563(29)$, $-0.101678(32)$, and $-0.145792(16)$ instead of $-2.410578(12)$, $-0.101676(3)$, and $-0.145792(2)$ for the $(1s1s)_0$, $(1s2p_{1/2})_0$, and $(1s2p_{3/2})_2$ states, respectively. This is still competitive with the data from Ref.~\cite{Yerokhin:2022:022815}. Since the proposed approach allows one to achieve good accuracy in calculations with relatively small values of $|\kappa|$, its application in the methods utilizing the finite-basis-set representation for the electron Green's function seems promising.


\section{Summary \label{sec:4}}

In the present work, the efficient and practical approach to accelerate the partial-wave-expansion convergence of two-electron self-energy contribution has been proposed. The approach is based on the method developed in Ref.~\cite{Sapirstein:2023:042804} for the first-order self-energy part of the Lamb shift. The modification of the standard procedure consists in subtracting the slowly-converging term and calculating it separately in momentum space in the closed form, without applying any expansion in partial waves. Special attention has been paid to the vertex diagram, which was the main source of the numerical uncertainty in the previous calculations.

Test calculations of the two-electron self-energy contribution to the binding energies of He-like ions for a number of nuclear charges, $Z=10$, 16, 24, and 32, and low-lying states, $(1s1s)_0$, $(1s2p_{1/2})_0$, and $(1s2p_{3/2})_2$, have been carried out. The considerable improvement of the partial-wave-series behavior is found compared to the behavior exhibited within the standard approach. The more accurate values, than those available in the literature, are obtained in the calculations with a relatively small number of partial waves considered. Therefore, the application of the worked out approach may, in particular, considerably expand the capabilities of the methods, which uses the finite-basis-set representations for the electron Green's functions. There are prospects also for further extension and development of the discussed approach.


\section*{Acknowledgments}

The authors are grateful to V. A. Yerokhin for sharing with us some details of the calculations from Ref.~\cite{Yerokhin:2022:022815}. The work was supported by the Russian Science Foundation (Grant No. 22-62-00004, https://rscf.ru/project/22-62-00004/). The calculations of the momentum-space contributions were supported by the Foundation for the Advancement of Theoretical Physics and Mathematics BASIS (Project No. 21-1-3-52-1).


\appendix


\section{Partial-wave expansion \label{sec:app:1}}

In the Feynman gauge, the photon propagator has the form
\begin{align}
\label{eq:A:D}
D_{\mu \nu}(\omega,\bx_{12}) = g_{\mu\nu} \, \frac{\exp\left[i\sqrt{\omega^2+i0}\,|\bx_{12}|\right]}{4\pi|\bx_{12}|} \,,
\end{align}
where the brunch of the square root is fixed with the condition ${\rm Im}(\sqrt{\omega^2+i0})>0$. The partial-wave expansion of the photon propagator arises from the standard expression, 
\begin{widetext}
\begin{align}
\label{eq:A:photon}
\frac{e^{i\omega x_{12}}}{x_{12}} = 4\pi i \omega
\sum_{L=0}^\infty \sum_{M=-L}^L  j_L(\omega x_<) h_L^{(1)}(\omega x_>) Y^*_{LM}(\bnx_1) Y_{LM}(\bnx_2) \, ,
\end{align}
where $j_L(z)$ and $h^{(1)}_L(z)$ are the spherical Bessel functions, $x_<={\rm min}(x_1,x_2)$, $x_>={\rm max}(x_1,x_2)$, and $\bnx = \bx/x$.

The partial-wave representation of the electron Green's function reads as
\begin{align}
\label{eq:A:G_expn}
G(E,\bx_1,\bx_2) = \sum_{\kappa} 
 \left( \begin{array}{rr} 
  G_{\kappa}^{11}(E,x_1,x_2) \pi^{++}_{\kappa}(\bnx_1,\bnx_2)
& 
-iG_{\kappa}^{12}(E,x_1,x_2) \pi^{+-}_{\kappa}(\bnx_1,\bnx_2)
\\[1mm]
 iG_{\kappa}^{21}(E,x_1,x_2) \pi^{-+}_{\kappa}(\bnx_1,\bnx_2)
& 
  G_{\kappa}^{22}(E,x_1,x_2) \pi^{--}_{\kappa}(\bnx_1,\bnx_2)
\\ \end{array} \right) \, ,
\end{align}
\end{widetext}
where $\pi^{\pm\pm}_{\kappa}(\bnx_1,\bnx_2)=\sum_{\mu} \Omega_{\pm\kappa\mu}(\bnx_1) \Omega^\dagger_{\pm\kappa\mu}(\bnx_2)$. Here, $\Omega_{\kappa\mu}$ is the spinor spherical harmonic~\cite{Varshalovich:1988:book:eng} and $\mu$ is the angular-momentum projection. The radial Green's function~$G_{\kappa}^{ik}$ can be constructed from the solutions of the radial Dirac equation bounded at infinity and at origin. For the free electron, these solutions can be expressed in terms of the spherical Bessel functions. In the case of the point-nucleus Coulomb potential, they can be written in terms of the Whittaker functions. We refer the reader, e.g., to Refs.~\cite{Mohr:1998:227, Yerokhin:2020:800} for further details.


\section{Momentum-space contributions arising from the free-electron self-energy operator \label{sec:app:2}}

The closed-form momentum-space expression for the subtraction in Fig.~\ref{fig:calc_4} is given in Eq.~(46) of Ref.~\cite{Sapirstein:2023:042804}, where the corresponding acceleration scheme was proposed. In this Appendix, we present the formula rewritten in the ``language'' of Refs.~\cite{Yerokhin:1999:800, Yerokhin:1999:3522} and in a form suitable for calculating the off-diagonal matrix elements of the SE operator~$\Sigma$ defined in Eq.~(\ref{eq:Sigma}). The derivation is based on the free-electron SE operator, which in the Feynman gauge is given by the integral
\begin{align}
\label{eq:A:SE_0_unR}
\Sigma^{(0)}({\rm p}) = -4\pi i \alpha 
\int\! \frac{d^4 {\rm k}}{(2\pi)^4} \, 
\frac{1}{{\rm k}^2}\gamma_\nu 
\frac{{\slashed{\rm p}}-{\slashed{\rm k}}+m}{({\slashed{\rm p}}-{\slashed{\rm k}})^2-m^2} \gamma^\nu \, .
\end{align} 
Here and below, the roman style is used for four-vectors, ${\rm k}=(k^0,{\bm k})$, the scalar product of two four vectors is ${\rm(pk)}=p^0k^0-\bp{\bm k}$, and ${\slashed{\rm k}}\equiv k_{\mu}\gamma^\mu$.

Let us first introduce some notations. The bound-electron solution of the Dirac equation~(\ref{eq:DirEq}) can be written in the form 
\begin{align}
\label{eq:A:wf_x}
\psi_a(\bx) = \dvec{g_a(x) \Omega_{\kappa_a\mu_a}(\bnx)}{i f_a(x) \Omega_{-\kappa_a\mu_a}(\bnx)} \, ,
\end{align}
where $g_a$ and $f_a$ are the large and small radial components. The Fourier transform of the coordinate-space wave function~(\ref{eq:A:wf_x}) leads to
\begin{align}
\label{eq:A:wf_p}
\psi_a(\bp) = \int \! d\bx \, e^{-i\bp\bx} \psi_a(\bx) = 
i^{-l_a}
\dvec{\tilde{g}_a(p) \Omega_{\kappa_a\mu_a}(\bnp)}{\tilde{f}_a(p) \Omega_{-\kappa_a\mu_a}(\bnp)} \, .
\end{align}
The first, i.e., zero-potential, term in Fig.~\ref{fig:calc_1} includes the renormalized free-electron SE operator~$\Sigma^{(0)}_{\rm R}$. In the off-diagonal case, this term reads as
\begin{align}
\label{eq:A:se_0}
E_{\rm 0P}&(\veps) = \frac{\alpha}{4\pi} \int_0^\infty \! \frac{dp\, p^2}{(2\pi)^3} \left\{ 
A(\rho)\left( \tilde{g}_a\tilde{g}_b - \tilde{f}_a\tilde{f}_b \right) \right. \nonumber \\
&\left.
+\, B(\rho) \left[ \veps \left( \tilde{g}_a\tilde{g}_b + \tilde{f}_a\tilde{f}_b \right) 
+ p \left( \tilde{g}_a\tilde{f}_b + \tilde{f}_a\tilde{g}_b \right)  \right]
\right\} \, ,
\end{align}
where the dependence of $\tilde{g}$ and $\tilde{f}$ on $p$ is omitted for brevity, and $\rho = 1 + (p^2-\veps^2)/m^2$, see Ref.~\cite{Yerokhin:1999:800} for details. In Eq.~(\ref{eq:A:se_0}), the energy~$\veps$ is the timelike component of the electron four-momentum, while $p$ is the magnitude of its spacelike component, that is ${\rm p}=(\veps,\bp)$ with $p=|\bp|$. According to the renormalization procedure, $\veps$ should be set to $\veps_a$ or $\veps_b$~\cite{TTGF}. However, we will leave it as a free parameter for a while. The other notations in Eq.~(\ref{eq:A:se_0}) are the following
\begin{align}
\label{eq:A:a}
A(\rho) &= 2m \left( 1 + \frac{2\rho}{1-\rho} \,{\rm ln}\,\rho \right) \, , \\
\label{eq:A:b}
B(\rho) &= -\frac{2-\rho}{1-\rho} \left( 1 + \frac{\rho}{1-\rho}\, {\rm ln}\,\rho \right) \, .
\end{align}

The subtraction in Fig.~\ref{fig:calc_4} approximates the contribution of the two-potential term $G^{(2)}=G^{(0)}VG^{(0)}VG^{(0)}$. The approximation consists in moving both potentials~$V$ out of the electron Green's function. In accordance with Eq.~(\ref{eq:GGG}), up to a factor of $1/2$, the resulting inner electron line represents $\partial^2 G^{(0)}(\veps-\omega,\bx_1,\bx_2)/\partial \veps^2$ sandwiched between two potentials. In the transition to momentum space, the potentials~$V(x)$ can be treated along with the wave functions~$\psi(\bx)$. Therefore, we define the Fourier transform of the wave function~(\ref{eq:A:wf_x}) multiplied by the potential via
\begin{align}
\label{eq:A:Vpsi_p}
\int \! d\bx \, e^{-i\bp\bx} V(x) \psi_a(\bx) = 
i^{-l_a}
\dvec{\tilde{t}_a(p) \Omega_{\kappa_a\mu_a}(\bnp)}{\tilde{s}_a(p) \Omega_{-\kappa_a\mu_a}(\bnp)} \, .
\end{align}
The momentum-space form of Eq.~(\ref{eq:GGG}) for the free-electron propagator is
\begin{align}
\label{eq:A:GGG_p}
\left(\frac{1}{{\slashed{\rm q}}-m+i0}\,\gamma^0\right)^3 = 
\frac{1}{2} \frac{\partial^2 }{\partial E^2} \,
\frac{1}{{\slashed{\rm q}}-m+i0}\,\gamma^0 \, ,
\end{align}
where ${\rm q}=(E,\bq)$. Putting it all together, one obtains that the subtraction can be expressed as
\begin{widetext}
\begin{align}
\label{eq:A:se_2P_approx}
\tilde{E}_{\rm 2P}&(\veps) = \frac{\alpha}{8\pi} \int_0^\infty \! \frac{dp\, p^2}{(2\pi)^3} \left\{ 
\frac{\partial^2 A}{\partial \veps^2} \left( \tilde{t}_a\tilde{t}_b - \tilde{s}_a\tilde{s}_b \right)
+ \left[ \veps \, \frac{\partial^2 B}{\partial \veps^2} + 2 \, \frac{\partial B}{\partial \veps} \right]
\left( \tilde{t}_a\tilde{t}_b + \tilde{s}_a\tilde{s}_b \right) 
+ p \,  \frac{\partial^2 B}{\partial \veps^2} \left( \tilde{t}_a\tilde{s}_b + \tilde{s}_a\tilde{t}_b \right)
\right\}  \, ,
\end{align}
where there is no need to keep the free parameter $\veps$ further, and one has to replace it with $\veps_a$ or $\veps_b$. The evaluation of the derivatives of Eqs.~(\ref{eq:A:a}) and (\ref{eq:A:b}) is straightforward,
\begin{align}
\label{eq:A:d2AdE2}
\frac{\partial^2 A}{\partial \veps^2} &= 
-\frac{8}{m(1-\rho)} \left\{ 
1 +
\frac{1}{1-\rho} 
\left[ 
{\rm ln}\,\rho  - \frac{2\veps^2}{m^2}
\left( 
\frac{1+\rho}{\rho} + \frac{2}{1-\rho}\, {\rm ln}\,\rho 
\right)
\right]
\right\} \, , \\
\label{eq:A:dBdE}
\frac{\partial B}{\partial \veps} &=  \frac{2\veps}{m^2(1-\rho)^2}
\left\{
3-\rho + \frac{2}{1-\rho} \, {\rm ln}\,\rho 
\right\} \, , \\
\label{eq:A:d2BdE2}
\frac{\partial^2 B}{\partial \veps^2} &= 
\frac{2}{m^2(1-\rho)^2}
\left\{
3-\rho + \frac{2}{1-\rho}
\left[
{\rm ln}\,\rho - \frac{\veps^2}{m^2}
\left( 
\frac{2+5\rho-\rho^2}{\rho} + \frac{6}{1-\rho} \, {\rm ln}\,\rho 
\right)
\right]
\right\} \, .
\end{align}
\end{widetext}
These derivatives are regular functions of $\rho$ at $\rho\approx 1$, that is at $p^2\approx \veps^2$. However, to avoid numerical problems, one can replace the exact expressions with their Taylor series in a small vicinity of this point. 

We note that in the case of the B-reducible term, the momentum-space expression for the subtraction can be readily obtained from Eq.~(\ref{eq:A:se_2P_approx}). The B-reducible term in Eq.~(\ref{eq:E_redB}) contains the derivative of the SE operator with respect to the argument. Its free part is derived from Eq.~(\ref{eq:A:se_0}) using the momentum-space form of Eq.~(\ref{eq:GG}), 
\begin{align}
\label{eq:A:GG_p}
\left(\frac{1}{{\slashed{\rm q}}-m+i0}\,\gamma^0\right)^2 = 
-\frac{\partial }{\partial E} \,
\frac{1}{{\slashed{\rm q}}-m+i0}\,\gamma^0 \, .
\end{align}
The scheme in Fig.~\ref{fig:calc_6}, applied to this contribution, implies one transposition of the potential~$V$, which can be treated by means of an additional differentiation. Therefore, to derive the desired formula from Eq.~(\ref{eq:A:se_2P_approx}), one has to: (i) restore one initial wave function (\ref{eq:A:wf_p}); (ii) change the overall sign to properly consider the differentiation in Eq.~(\ref{eq:E_redB}); (iii) take into account the factor associated with the matrix element of the operator $I$. Additionally, the resulting expression can be multiplied by two in order to account for that both subtractions in Fig.~\ref{fig:calc_6} coincide in this case.


\section{Momentum-space contributions arising from the free-electron vertex operator \label{sec:app:3}}

The subtractions in Figs.~\ref{fig:calc_5} and \ref{fig:calc_6} involve only one transposition of the potential~$V$ from an inner electron line to the nearest vertex, where the photon is connected. The role of a generating expression in this case is played by the free-electron vertex operator,
\begin{widetext}
\begin{align}
\label{eq:A:Gamma}
\Gamma^{\mu}({\rm p},{\rm p}') = -4\pi i \alpha \int \! \frac{d^4 {\rm k}}{(2\pi)^4} \,
\frac{1}{{\rm k}^2} \gamma_\nu
\frac{{\slashed{\rm p}}-{\slashed{\rm k}}+m}{({\rm p}-{\rm k})^2-m^2} \gamma^\mu 
\frac{{\slashed{\rm p}}'-{\slashed{\rm k}}+m}{({\rm p}'-{\rm k})^2-m^2} \gamma^\sigma \,.
\end{align}
To make the discussion complete and to unify the notations used, in this Appendix we compile all the relevant formulas, some of which were previously given in Refs.~\cite{Yerokhin:1999:800, Yerokhin:1999:3522}.

The second, i.e., one-potential, term in Fig.~\ref{fig:calc_1} and the first, i.e., free-vertex, term in Fig.~\ref{fig:calc_3} can be written as
\begin{align}
\label{eq:A:se_1}
E_{\rm 1P}(\veps,\veps') &= \int \! \frac{d \bp}{(2\pi)^3} \int \! \frac{d \bp'}{(2\pi)^3} \,
\bar{\psi}_a(\bp) V(|\bq|) \Gamma_{\rm R}^0({\rm p},{\rm p}') \psi_b(\bp') \, ,  \\
\label{eq:A:vert_0}
E_{\rm vert}^{(0)}(\veps,\veps') &= \int \! \frac{d \bp}{(2\pi)^3} \int \! \frac{d \bp'}{(2\pi)^3} \,
\bar{\psi}_a(\bp) A_{\mu}^{cd}(\bq) \Gamma_{\rm R}^{\mu}({\rm p},{\rm p}') \psi_b(\bp') \, ,
\end{align}
where, $\bar{\psi}=\psi^\dagger\gamma^0$, $\bq=\bp-\bp'$, $V(|\bq|)$ is the Fourier transform of the spherically symmetric potential $V(x)$, and
\begin{align}
A_{\mu}^{cd}(\bq) = \frac{4\pi\alpha}{\bq^2 -\Delta_{d,c}^2-i0} \, 
\int \! d\bz \, e^{-i\bq\bz} \psi^\dagger_c(\bz) \alpha_\mu \psi_d(\bz) \, .
\end{align}
The free-vertex contribution in Eq.~(\ref{eq:A:vert_0}) is given not for the Slater-determinant state~(\ref{eq:u_2el}), but for a matrix element between two arbitrary two-electron wave functions $\psi_a\psi_c$ and $\psi_b\psi_d$ with the SE loop attributed to the ``$ab$'' electron line. The corresponding expression can be readily employed for any case of interest. The parameters $\veps$ and $\veps'$ in Eqs.~(\ref{eq:A:se_1}) and (\ref{eq:A:vert_0}) play a similar role to the variable $\veps$ in Eq.~(\ref{eq:A:se_0}), that is ${\rm p}=(\veps,\bp)$ and ${\rm p}'=(\veps',\bp')$. We keep them as free parameters, but at the final stage in Eq.~(\ref{eq:A:se_1}) they should both be set to $\veps_a$ or $\veps_b$, while in Eq.~(\ref{eq:A:vert_0}) one must replace $\veps$ and $\veps'$ with $\veps_a$ and $\veps_b$, respectively. 

According to Ref.~\cite{Yerokhin:1999:800}, the renormalized free-electron vertex operator is given by
\begin{align}
\label{eq:A:Gamma_R}
\Gamma_{\rm R}^{\mu}({\rm p},{\rm p}') 
&= \frac{\alpha}{4\pi} \left\{
A \gamma^\mu +  {\slashed{\rm p}} \left( B_1 p^{\mu} + B_2 p^{\prime\mu} \right)
+ {\slashed{\rm p}}' \left( C_1 p^{\mu} + C_2 p^{\prime\mu} \right)
+ D \left( {\slashed{\rm p}} \gamma^\mu {\slashed{\rm p}}' \right) + H_1 p^{\mu} + H_2 p^{\prime\mu} 
\right\} \, , \\
\label{eq:A:coef_A}
A&=C_{24} - 2 + {\rm p}^2 C_{11} + {\rm p}^{\prime 2} C_{12}
+ 4({\rm p}{\rm p}')\left( C_0 + C_{11} + C_{12} \right)
+ m^2 \left( -2\,C_0 + C_{11} + C_{12} \right) \, , 
\end{align}
\end{widetext}
\begin{align}
\label{eq:A:coef_B1}
B_1 &= -4\left(C_{11} +C_{21}\right) \, , \\
\label{eq:A:coef_B2}
B_2 &= -4 \left(C_0 + C_{11} + C_{12} + C_{23}\right) \, , \\
\label{eq:A:coef_C1}
C_1 &= -4 \left(C_0 + C_{11} + C_{12} + C_{23}\right) \, , \\
\label{eq:A:coef_C2}
C_2 &= -4\left(C_{12} +C_{22}\right) \, , \\
\label{eq:A:coef_D}
D &= 2\left( C_0 + C_{11} + C_{12} \right) \, , \\
\label{eq:A:coef_H1} 
H_1 &= 4m \left( C_0 + 2\,C_{11} \right) \, , \\
\label{eq:A:coef_H2} 
H_2 &= 4m \left( C_0 + 2\,C_{12} \right) \, . 
\end{align}
The coefficients $C_0$ and $C_{ij}$ in Eqs.~(\ref{eq:A:coef_A})-(\ref{eq:A:coef_H2}) are defined by
\begin{align}
\label{eq:A:C_0}
C_0 &= \int_0^1 \frac{dy}{u}\big[ -{\rm ln}(1+\beta) \big] \, , \\
\label{eq:A:C_1i}
\dvec{C_{11}}{C_{12}} &= \int_0^1 \frac{dy}{u} \dvec{y}{1-y}
\left[ 1 - \frac{{\rm ln}(1+\beta)}{\beta} \right] \, , \\
\label{eq:A:C_2i}
\begin{pmatrix} C_{21} \\ C_{22} \\ C_{23} \end{pmatrix} &=
\int_0^1 \frac{dy}{u} 
\begin{pmatrix} y^2 \\ (1-y)^2 \\ y(1-y) \end{pmatrix}
\left[ -\frac{1}{2} + \frac{1}{\beta} - \frac{{\rm ln}(1+\beta)}{\beta^2} \right] \, , \\
\label{eq:A:C_24}
C_{24} &= - \int_0^1 \! dy \, {\rm ln}\!
\left( y(y-1) \frac{\rm q^2}{m^2} + 1 \right) \, ,
\end{align}
where $\rm q = p - p'$, $\beta = u / v$, and
\begin{align}
\label{eq:A:b}
u &= \left( y {\rm p} + (1-y) {\rm p}' \right)^2 \, , \\
\label{eq:A:s}
v &= m^2 -y {\rm p}^2 - (1-y) {\rm p}^{\prime 2} \, .
\end{align} 
To perform the angular integrations in Eqs.~(\ref{eq:A:se_1}) and (\ref{eq:A:vert_0}), it is convenient to use the following expressions for the timelike, $\Gamma_{\rm R}^0$, and spacelike, $\bm \Gamma_{\rm R}$, components of the free-electron vertex operator sandwiched between two Dirac wave functions, 
\begin{widetext}
\begin{align}
\label{eq:A:psiG0psi}
\bar{\psi}_a(\bp) \Gamma_{\rm R}^0({\rm p},{\rm p}') \psi_b(\bp') &=
\frac{\alpha}{4\pi} \, i^{l_a-l_b} 
\left\{
\mathcal{F}_1^{a,b} \, \Omega^\dagger_{\kappa_a\mu_a}(\bnp)  \Omega_{\kappa_b\mu_b}(\bnp')
+
\mathcal{F}_2^{a,b} \, \Omega^\dagger_{-\kappa_a\mu_a}(\bnp) \Omega_{-\kappa_b\mu_b}(\bnp') 
\right\} \, , \\
\label{eq:A:psiGipsi}
\bar{\psi}_a(\bp) {\bm \Gamma_{\rm R}}({\rm p},{\rm p}') \psi_b(\bp') &=
\frac{\alpha}{4\pi} \, i^{l_a-l_b} 
\left\{
\mathcal{R}_1^{a,b} \, \Omega^\dagger_{\kappa_a\mu_a}(\bnp) \bsigma \Omega_{-\kappa_b\mu_b}(\bnp')
+
\mathcal{R}_2^{a,b} \, \Omega^\dagger_{-\kappa_a\mu_a}(\bnp) \bsigma \Omega_{\kappa_b\mu_b}(\bnp') \right. \nonumber \\
&\quad+\left.
\left( \mathcal{R}_3^{a,b} \bp + \mathcal{R}_4^{a,b} \bp' \right) 
\Omega^\dagger_{\kappa_a\mu_a}(\bnp)  \Omega_{\kappa_b\mu_b}(\bnp')
+
\left( \mathcal{R}_5^{a,b} \bp + \mathcal{R}_6^{a,b} \bp' \right) 
\Omega^\dagger_{-\kappa_a\mu_a}(\bnp) \Omega_{-\kappa_b\mu_b}(\bnp') 
\right\} \, .
\end{align}
Here $\bsigma$ is the vector of the Pauli matrices and the coefficients $\mathcal{F}_i^{a,b}$ and $\mathcal{R}_i^{a,b}$ depend on $p=|\bp|$, $p'=|\bp'|$, and $\xi=\bnp\bnp'$, which is the cosine of the angle between $\bp$ and $\bp'$. The dependence of $\mathcal{F}_i^{a,b}$ and $\mathcal{R}_i^{a,b}$ on the parameters~$\veps$ and $\veps'$ is also implied. These coefficients are defined as follows:
\begin{align}
\label{eq:A:F1}
\mathcal{F}_1^{a,b} &= 
\left( A + H_1 \veps + H_2 \veps' \right) \tilde{g}_a \tilde{g}'_b
+ \left( B_1 \veps + B_2 \veps' \right) \left( \veps \tilde{g}_a + p \tilde{f}_a \right) \tilde{g}'_b \nonumber \\
&+ \left( C_1 \veps + C_2 \veps' \right) \tilde{g}_a \left( \veps' \tilde{g}'_b + p' \tilde{f}'_b \right)  
+ D \left( \veps \tilde{g}_a + p \tilde{f}_a \right) \left( \veps' \tilde{g}'_b + p' \tilde{f}'_b \right) \, , \\
\label{eq:A:F2}
\mathcal{F}_2^{a,b} &= 
\left( A - H_1 \veps - H_2 \veps' \right) \tilde{f}_a \tilde{f}'_b
+ \left( B_1 \veps + B_2 \veps' \right) \left( \veps \tilde{f}_a + p \tilde{g}_a \right) \tilde{f}'_b  \nonumber \\
&+ \left( C_1 \veps + C_2 \veps' \right) \tilde{f}_a \left( \veps' \tilde{f}'_b + p' \tilde{g}'_b \right) 
+ D \left( \veps \tilde{f}_a + p \tilde{g}_a \right) \left( \veps' \tilde{f}'_b + p' \tilde{g}'_b \right) \, , \\
\label{eq:A:R1}
\mathcal{R}_1^{a,b} &= 
A \tilde{g}_a \tilde{f}'_b
- D \left( \veps \tilde{g}_a + p \tilde{f}_a \right) \left( \veps' \tilde{f}'_b + p' \tilde{g}'_b \right) \, , \\
\label{eq:A:R2}
\mathcal{R}_2^{a,b} &= 
A \tilde{f}_a \tilde{g}'_b
- D \left( \veps \tilde{f}_a + p \tilde{g}_a \right) \left( \veps' \tilde{g}'_b + p' \tilde{f}'_b \right) \, , \\
\label{eq:A:R3}
\mathcal{R}_3^{a,b} &=
B_1 \left( \veps \tilde{g}_a + p \tilde{f}_a \right) \tilde{g}'_b
+ C_1 \tilde{g}_a \left( \veps' \tilde{g}'_b + p' \tilde{f}'_b \right)
+ H_1 \tilde{g}_a \tilde{g}'_b \, , \\
\label{eq:A:R4}
\mathcal{R}_4^{a,b} &=
B_2 \left( \veps \tilde{g}_a + p \tilde{f}_a \right) \tilde{g}'_b
+ C_2 \tilde{g}_a \left( \veps' \tilde{g}'_b + p' \tilde{f}'_b \right)
+ H_2 \tilde{g}_a \tilde{g}'_b \, , \\
\label{eq:A:R5}
\mathcal{R}_5^{a,b} &=
B_1 \left( \veps \tilde{f}_a + p \tilde{g}_a \right) \tilde{f}'_b
+ C_1 \tilde{f}_a \left( \veps' \tilde{f}'_b + p' \tilde{g}'_b \right)
- H_1 \tilde{f}_a \tilde{f}'_b \, , \\
\label{eq:A:R6}
\mathcal{R}_6^{a,b} &=
B_2 \left( \veps \tilde{f}_a + p \tilde{g}_a \right) \tilde{f}'_b
+ C_2 \tilde{f}_a \left( \veps' \tilde{f}'_b + p' \tilde{g}'_b \right)
- H_2 \tilde{f}_a \tilde{f}'_b \, .
\end{align}
For brevity, the dependence of wave functions on $p$ and $p'$ is omitted. For the functions of $p'$, an additional prime is added. Therefore, the shorthand notations are $\tilde{g}_a = \tilde{g}_a(p)$, $\tilde{f}_a = \tilde{f}_a(p)$, $\tilde{g}'_b = \tilde{g}_b(p')$, and $\tilde{f}'_b = \tilde{f}_b(p')$. After the angular integration is performed, the one-potential term~(\ref{eq:A:se_1}) reads
\begin{align}
\label{eq:A:se_1_}
E_{\rm 1P}(\veps,\veps') &= \frac{\alpha}{4\pi} \frac{1}{(2\pi)^5} \, 
\int_0^\infty \! dp \int_0^\infty \! dp' \int_{-1}^1 \! d \xi \, p^2 p^{\prime 2} \, V(q)
\left\{
\mathcal{F}_1^{a,b}(p,p',\xi) P_l(\xi) + \mathcal{F}_2^{a,b}(p,p',\xi) P_{\bar l}(\xi)
\right\} \, , 
\end{align}
\end{widetext}
where $q^2=p^2+p^{\prime 2}-2pp'\xi$, $l=|\kappa_a+1/2|-1/2$, $\bar l = 2j-l$, $j=|\kappa_a|-1/2$, $P_l$ is the Legendre polynomial, and $\kappa_a=\kappa_b$ due to the conservation of angular quantum numbers by the SE operator~\cite{Yerokhin:1999:800}. The angular integration for the free-vertex contribution~(\ref{eq:A:vert_0}) is a bit more complicated. Since the angular parts of Eq.~(\ref{eq:A:vert_0}) and the desired subtraction are the same, we do not present the corresponding formulas here and refer the reader to Ref.~\cite{Yerokhin:1999:3522} for details. Using the expressions given below, the standard numerical code for calculating the free-vertex contribution can be readily modified to handle the subtraction in momentum space.

A general roadmap for deriving the closed-form momentum-space expression for the subtractions in Figs.~\ref{fig:calc_5} and \ref{fig:calc_6} is as follows: (i) differentiate the ``base'' expressions~(\ref{eq:A:se_1}) and (\ref{eq:A:vert_0}) with respect to $\veps$ or $\veps'$ depending on the electron line from which the potential~$V$ is transposed to the nearest vertex; (ii) replace the appropriate wave function (\ref{eq:A:wf_p}) with (\ref{eq:A:Vpsi_p}); (iii) change the overall sign according to Eq.~(\ref{eq:A:GG_p}). In this Appendix, we consider the differentiation with respect to $\veps$. The case of $\veps'$ is treated similarly.

According to Eqs.~(\ref{eq:A:psiG0psi}) and (\ref{eq:A:psiGipsi}), the differentiation of Eqs.~(\ref{eq:A:se_1}) and (\ref{eq:A:vert_0}) is equivalent to the differentiation of the coefficients $\mathcal{F}_i^{a,b}$ and $\mathcal{R}_i^{a,b}$. Differentiating Eqs.~(\ref{eq:A:F1})-(\ref{eq:A:R6}) with respect to $\veps$ and at the same time replacing the wave function (\ref{eq:A:wf_p}) for the state $|a\rangle$ by (\ref{eq:A:Vpsi_p}) gives:
\begin{widetext}
\begin{align}
\label{eq:A:dF1}
\frac{d\mathcal{F}_1^{V\!a,b}}{d\veps} &=
\left( \frac{dA}{d\veps} + B_1 \veps + B_2 \veps' 
     + H_1 + \frac{dH_1}{d\veps}\,\veps + \frac{dH_2}{d\veps}\,\veps' \right) \tilde{t}_a \tilde{g}'_b
+ \left( B_1 + \frac{dB_1}{d\veps}\,\veps + \frac{dB_2}{d\veps}\,\veps' \right) 
  \left( \veps \tilde{t}_a + p \tilde{s}_a \right) \tilde{g}'_b  \nonumber \\
&
+  \left( C_1 + \frac{dC_1}{d\veps}\,\veps + \frac{dC_2}{d\veps}\,\veps' + D \right) 
   \tilde{t}_a \left( \veps' \tilde{g}'_b + p' \tilde{f}'_b \right) 
+ \frac{dD}{d\veps} \left( \veps \tilde{t}_a + p \tilde{s}_a \right) \left( \veps' \tilde{g}'_b + p' \tilde{f}'_b \right) 
\, , \\
\label{eq:A:dF2}
\frac{d\mathcal{F}_2^{V\!a,b}}{d\veps} &=
\left( \frac{dA}{d\veps} + B_1 \veps + B_2 \veps' 
     - H_1 - \frac{dH_1}{d\veps}\,\veps - \frac{dH_2}{d\veps}\,\veps' \right) \tilde{s}_a \tilde{f}'_b
+ \left( B_1 + \frac{dB_1}{d\veps}\,\veps + \frac{dB_2}{d\veps}\,\veps' \right) 
  \left( \veps \tilde{s}_a + p \tilde{t}_a \right) \tilde{f}'_b  \nonumber \\
&
+  \left( C_1 + \frac{dC_1}{d\veps}\,\veps + \frac{dC_2}{d\veps}\,\veps' + D \right) 
   \tilde{s}_a \left( \veps' \tilde{f}'_b + p' \tilde{g}'_b \right) 
+ \frac{dD}{d\veps} \left( \veps \tilde{s}_a + p \tilde{t}_a \right) \left( \veps' \tilde{f}'_b + p' \tilde{g}'_b \right)
\, , \\
\label{eq:A:dR1}
\frac{d\mathcal{R}_1^{V\!a,b}}{d\veps} &=
\frac{dA}{d\veps} \, \tilde{t}_a \tilde{f}'_b
- \frac{dD}{d\veps} \left( \veps \tilde{t}_a + p \tilde{s}_a \right) \left( \veps' \tilde{f}'_b + p' \tilde{g}'_b \right)
- D \, \tilde{t}_a \left( \veps' \tilde{f}'_b + p' \tilde{g}'_b \right) \, , \\
\label{eq:A:dR2}
\frac{d\mathcal{R}_2^{V\!a,b}}{d\veps} &=
\frac{dA}{d\veps} \, \tilde{s}_a \tilde{g}'_b
- \frac{dD}{d\veps} \left( \veps \tilde{s}_a + p \tilde{t}_a \right) \left( \veps' \tilde{g}'_b + p' \tilde{f}'_b \right)
- D \, \tilde{s}_a \left( \veps' \tilde{g}'_b + p' \tilde{f}'_b \right) \, , \\
\label{eq:A:dR3}
\frac{d\mathcal{R}_3^{V\!a,b}}{d\veps} &=
\left( B_1 + \frac{dH_1}{d\veps} \right) \tilde{t}_a \tilde{g}'_b
+ \frac{dB_1}{d\veps} \left( \veps \tilde{t}_a + p \tilde{s}_a \right) \tilde{g}'_b
+ \frac{dC_1}{d\veps} \, \tilde{t}_a \left( \veps' \tilde{g}'_b + p' \tilde{f}'_b \right) \, , \\
\label{eq:A:dR4}
\frac{d\mathcal{R}_4^{V\!a,b}}{d\veps} &=
\left( B_2 + \frac{dH_2}{d\veps} \right) \tilde{t}_a \tilde{g}'_b
+ \frac{dB_2}{d\veps} \left( \veps \tilde{t}_a + p \tilde{s}_a \right) \tilde{g}'_b
+ \frac{dC_2}{d\veps} \, \tilde{t}_a \left( \veps' \tilde{g}'_b + p' \tilde{f}'_b \right) \, , \\
\label{eq:A:dR5}
\frac{d\mathcal{R}_5^{V\!a,b}}{d\veps} &=
\left( B_1 - \frac{dH_1}{d\veps} \right) \tilde{s}_a \tilde{f}'_b
+ \frac{dB_1}{d\veps} \left( \veps \tilde{s}_a + p \tilde{t}_a \right) \tilde{f}'_b
+ \frac{dC_1}{d\veps} \, \tilde{s}_a \left( \veps' \tilde{f}'_b + p' \tilde{g}'_b \right) \, , \\
\label{eq:A:dR6}
\frac{d\mathcal{R}_6^{V\!a,b}}{d\veps} &=
\left( B_2 - \frac{dH_2}{d\veps} \right) \tilde{s}_a \tilde{f}'_b
+ \frac{dB_2}{d\veps} \left( \veps \tilde{s}_a + p \tilde{t}_a \right) \tilde{f}'_b
+ \frac{dC_2}{d\veps} \, \tilde{s}_a \left( \veps' \tilde{f}'_b + p' \tilde{g}'_b \right) \, .
\end{align}
In Eqs.~(\ref{eq:A:dF1})-(\ref{eq:A:dR6}), the index $V$ is added to emphasize that the wave function of the state $|a\rangle$ is multiplied by the potential. Taking into account that ${\rm p}^2=\veps^2 - p^2$ and $({\rm p}{\rm p'})=\veps\veps'-pp'\xi$, one obtains
\begin{align}
\label{eq:A:dA}
\frac{dA}{d\veps} &= \frac{dC_{24}}{d\veps} + 2\veps \, C_{11}
+ {\rm p}^2 \,\frac{dC_{11}}{d\veps} + {\rm p}^{\prime 2} \,\frac{dC_{12}}{d\veps}
+ 4 \veps' \left( C_0 + C_{11} + C_{12} \right) \nonumber \\
&+ 4({\rm p}{\rm p}')\left( \frac{dC_0}{d\veps} + \frac{dC_{11}}{d\veps} + \frac{dC_{12}}{d\veps} \right)
+ m^2 \left( -2\,\frac{dC_0}{d\veps} + \frac{dC_{11}}{d\veps} + \frac{dC_{12}}{d\veps} \right) \, .
\end{align}
Since Eqs.~(\ref{eq:A:coef_B1})-(\ref{eq:A:coef_H2}) are linear combinations of the coefficients $C_0$ and $C_{ij}$, their differentiation is trivial. Therefore, we proceed to the differentiation of the coefficients themselves. Differentiating Eqs.~(\ref{eq:A:C_0})-(\ref{eq:A:C_2i}) with respect to $\veps$ yields
\begin{align}
\label{eq:A:dC_0}
\frac{dC_0}{d\veps} &=
\int_0^1 \frac{dy}{u} 
\left\{ 
\frac{1}{v}\frac{dv}{d\veps} \, {\rm ln}(1+\beta) 
+ 
\frac{1}{\beta} \frac{d\beta}{d\veps} 
\left[ {\rm ln}(1+\beta) - \frac{\beta}{1+\beta}\right]
\right\} \, , \\
\label{eq:A:dC_1i}
\frac{d}{d\veps} \dvec{C_{11}}{C_{12}}  &=\int_0^1 \frac{dy}{u} \dvec{y}{1-y}
\left\{
\frac{1}{v}\frac{dv}{d\veps} \left[ \frac{{\rm ln}(1+\beta)}{\beta} - 1 \right]
+
\frac{1}{\beta} \frac{d\beta}{d\veps}
\left[ \frac{2\,{\rm ln}(1+\beta)}{\beta} - \frac{2+\beta}{1+\beta} \right]
\right\} \, , \\
\label{eq:A:dC_2i}
\frac{d}{d\veps}  \begin{pmatrix} C_{21} \\ C_{22} \\ C_{23} \end{pmatrix} &=
\int_0^1 \frac{dy}{u} \begin{pmatrix} y^2 \\ (1-y)^2 \\ y(1-y) \end{pmatrix}
\left\{
\frac{1}{v}\frac{dv}{d\veps} \left[ \frac{{\rm ln}(1+\beta)}{\beta^2} - \frac{1}{\beta} + \frac{1}{2} \right]
+
\frac{1}{\beta} \frac{d\beta}{d\veps}
\left[
\frac{3\,{\rm ln}(1+\beta)}{\beta^2} - \frac{3+2\beta}{\beta(1+\beta)} + \frac{1}{2}
\right]
\right\} \, ,
\end{align}
\end{widetext}
where
\begin{align}
\label{eq:A:dbeta}
\frac{d\beta}{d\veps} &= \frac{d}{d\veps} \bigg( \frac{u}{v} \bigg)
= \frac{1}{v}\left[ \frac{du}{d\veps} - \beta \frac{dv}{d\veps} \right] \, , \\
\label{eq:A:du}
\frac{du}{d\veps} & = 2y \left[ y \veps + (1-y) \veps' \right] \, , \\
\label{eq:A:dv}
\frac{dv}{d\veps} & = - 2y \veps \, .
\end{align}
Finally, for the coefficient~$C_{24}$, we have
\begin{align}
\label{eq:A:dC24}
\frac{dC_{24}}{d\veps} &= - \int_0^1 \! dy \, 
\frac{y(y-1)}{y(y-1) {\rm q^2} + m^2} \dfrac{d{\rm q^2}}{d\veps} \, ,
\end{align}
where $d{\rm q^2}/d\veps = 2(\veps-\veps')$.

Summing up all that has been discussed in this Appendix, the closed-form momentum-space expression for the subtraction in Fig.~\ref{fig:calc_5} reads as
\begin{widetext}
\begin{align}
\label{eq:A:se_2P_approx_2}
\tilde{E}_{\rm 2P}(\veps,\veps') &= - \frac{\alpha}{4\pi} \frac{1}{(2\pi)^5} \, 
\int_0^\infty \! dp \int_0^\infty \! dp' \int_{-1}^1 \! d \xi \, p^2 p^{\prime 2} \, V(q)
\left\{
\frac{d\mathcal{F}_1^{V\!a,b}(p,p',\xi)}{d\veps} P_l(\xi) + \frac{d\mathcal{F}_2^{V\!a,b}(p,p',\xi)}{d\veps} P_{\bar l}(\xi)
\right\} \, .
\end{align}
\end{widetext}
There is no need to keep the free parameters $\veps$ and $\veps'$ further, and one has to set $\veps=\veps'$ and replace them with $\veps_a$ or $\veps_b$. The expression for the subtraction in Fig.~\ref{fig:calc_6} can be obtained in a similar way. 

We note that the formulas derived here turn out to be more complicated than Eq.~(\ref{eq:A:se_2P_approx}). However, their complexity is comparable to that of the contributions which are to be evaluated during the renormalization procedure.




\end{document}